\documentclass[twocolumn,preprintnumbers,amssymb,amsmath,superscriptaddress]{revtex4}
\usepackage[dvips]{graphicx}
\usepackage{dcolumn} 
\usepackage{bm}
\usepackage{epsfig,amsmath,amssymb,verbatim,mathrsfs,array,layout,textcomp,amssymb,latexsym,slashed}
\usepackage{color}
\input epsf.tex
\newcommand{\beq}{\begin{eqnarray}}
\newcommand{\eeq}{\end{eqnarray}}
\newcommand{\nn}{\nonumber}
\def\ltap{\ \raise.3ex\hbox{$<$\kern-.75em\lower1ex\hbox{$\sim$}}\ }
\def\gtap{\ \raise.3ex\hbox{$>$\kern-.75em\lower1ex\hbox{$\sim$}}\ }

\def\be{\begin{equation}}
\def\ee{\end{equation}}
\def\bea{\begin{eqnarray}}
\def\eea{\end{eqnarray}}

\pdfoutput=1

\newcommand{\hc}{{\rm h.c.}}

\begin{document}

\title{
A Dark Matter Relic from Muon Anomalies
}
\author{Genevi\`eve B\'elanger}
\affiliation{LAPTh, Universit\'e Savoie Mont Blanc, CNRS B.P. 110, F-74941 Annecy-le-Vieux, France}
\author{C\'edric Delaunay}
\affiliation{LAPTh, Universit\'e Savoie Mont Blanc, CNRS B.P. 110, F-74941 Annecy-le-Vieux, France}
\author{Susanne Westhoff\,}
\affiliation{PITTsburgh Particle-physics Astrophysics \& Cosmology Center (PITT-PACC), Department of Physics \& Astronomy, University of Pittsburgh, Pittsburgh, PA 15260, USA}
\preprint{\scriptsize LAPTH-039/15, PITT-PACC-1511\vspace*{.1cm}\\}
\vskip .05in

\begin{abstract}
\vskip .05in
We show that the recently reported anomalies in $b\to s\mu^+\mu^-$ transitions, as well as the long-standing $g_\mu-2$ discrepancy, can be addressed simultaneously by a new  massive abelian gauge boson with loop-induced coupling to muons. Such a scenario typically leads to a stable dark matter candidate with a thermal relic density
 close to the observed value. Dark matter in our model couples dominantly to leptons, hence
 signals in direct detection experiments lie well below the current sensitivity.
  The LHC, in combination with indirect detection searches, can test this scenario through distinctive signatures with muon pairs and missing energy. 
\end{abstract}

\maketitle
\section{Introduction}\label{intro}
The lack of an acceptable Dark Matter (DM) candidate within the Standard Model (SM) is a pressing phenomenological motivation for the existence of new physics (NP) beyond the SM. 
 Dark Matter may very well be close to the weak scale, emerging from 
 theories addressing the electroweak (EW) hierarchy problem of the SM. 
However, these theories generally
 predict a variety of new phenomena and particles around the TeV scale, whose existence still remains to be established experimentally. 
Despite the lack of a NP discovery, a few measurements are in mild tensions with SM predictions.
 In particular, there is a growing array of anomalies involving muons~\cite{gm2BNL,Kstar,Kstarnew,LHCbRK}\footnote{We do not consider the proton size anomaly observed in muonic hydrogen atoms~\cite{muHnature}. NP interpretations of this measurement are very challenging~\cite{psizeBarger,MeVforces}.}.
 Moreover, DM searches at direct detection experiments strongly limit its interactions with quarks (see {\it e.g.} REF.~\cite{Akerib:2013tjd}), suggesting that
 DM around the EW scale might preferentially couple to leptons. In this paper, we thus
 take a data-driven approach, entertaining the possibility that the observed muon-related anomalies are the first signals of leptophilic DM at the EW scale.

The list of anomalies involving muons starts with the long-standing puzzle of the  anomalous magnetic moment of the muon, $a_\mu\equiv (g_\mu-2)/2$.
 The BNL measurement~\cite{gm2BNL} exceeds the SM prediction
 by about $3$ standard deviations~\cite{PDG},
\beq\label{gm2data}
\Delta a_\mu \equiv a_\mu^{\rm exp} -a^{\rm SM}_\mu = (287\pm80)\times 10^{-11}\,.
\eeq
More recently, the LHCb experiment reported on a series of anomalies in (semi-) leptonic $B$ meson decays, which together point to a possible new source of $b\to s \mu^+\mu^-$ transitions at short distances. The perhaps most tantalizing deviation occurs in the ratio $R_K=\mathcal{B}(B^+\to K^+\mu^+\mu^-)/\mathcal{B}(B^+\to K^+e^+e^-)$, which was observed about $2.6\,\sigma$  below
 the theoretically clean SM prediction, $R_K^{\rm SM}-1\sim 10^{-4}$~\cite{LHCbRK},
\beq
R_K^{\rm exp} =0.745^{+0.090}_{-0.074}({\rm stat})\pm0.036({\rm syst}).
\eeq
Furthermore,
 the measured decay rate for $B^0\to K^{0*}\mu^+\mu^-$
 was found to
exceed the SM prediction in a particular region of phase space~\cite{Kstar}, a result supported by the latest LHCb data~\cite{Kstarnew}. The observed decay rate for $B^0_s\to \mu^+\mu^-$ is
 slightly below, but compatible with the SM prediction~\cite{CMS:2014xfa}. 
 While none of these LHCb anomalies by themselves are significant enough to claim a discovery,
 it is intriguing that they point to a common Lorentz structure when interpreted as a signal of NP~\cite{Descotes-Genon:2013wba,Altmannshofer-Straub,Beaujean:2013soa,Altmannshofer:2014rta,Hurth:2014vma}. A simultaneous explanation of $\Delta a_{\mu}$, however, requires more sophisticated  model building. In this paper, we propose a simple
 toy model, which addresses both the $g_\mu-2$ and the various LHCb anomalies.

Most NP interpretations of the
$b\to s \mu^+\mu^-$ anomalies 
 postulate the existence of  a new state in the range of $\Lambda\sim 1-10$~TeV with {\it tree-level} couplings to muons and quarks, for instance a $Z'$ gauge boson~\cite{Descotes-Genon:2013wba,Gauld:2013qba,Altmannshofer:2014cfa,Sierra:2015fma} or a scalar leptoquark~\cite{Hiller:2014yaa,Biswas:2014gga,Gripaios:2014tna}.
 The same interactions typically contribute to $g_\mu-2$ at the one-loop level, but yield too small a contribution to explain the discrepancy in Eq.~\eqref{gm2data}, due to the suppression by the high scale, $a_{\mu}\propto m_\mu^2/\Lambda^2$~\cite{Altmannshofer-Straub}.
  A way out is to  generate the coupling of $Z'$ to muon pairs
 only {\it radiatively}, so that contributions to $b\to s\mu^+\mu^-$ transitions and $g_{\mu}-2$ are both induced at the one-loop level by NP around the EW scale.
 As we argue in this work, this requires a richer
 NP sector with an 
 electrically neutral state, which is
 stable if the tree-level $Z'\mu^+ \mu^-$ coupling is forbidden by a (spontaneously broken) symmetry. Hence, addressing the aforementioned muon-related anomalies generally
 {\it yields} a DM candidate, which, by construction, is mostly leptophilic.
 The same NP interactions dominate DM annihilation in the early universe.
We will demonstrate that a
 minimal
 model with the above properties typically
leads to a stable DM candidate with a
 thermal
 relic density of the order of the 
 observed value.
 
 The remainder of the paper is organized as follows. We introduce
 our phenomenological model in SEC.~\ref{model} and identify the
 parameter space to
 simultaneously accommodate the $g_\mu-2$ and $b\to s\mu^+\mu^-$ anomalies in SEC.~\ref{sec:anomalies}.  We then discuss the implications of 
 collider constraints in SEC.~\ref{collconst}. The resulting DM phenomenology 
 is analysed in SEC.~\ref{DM} and SEC.~\ref{ID}. We conclude and give an outlook on future experimental tests of our model
 in SEC.~\ref{CO}.  

\section{A Minimal Model}\label{model}
We consider  the extension of the SM
 by a new (dark) sector, consisting of heavy leptons $L$, $L^c$ and quarks $Q$, $Q^c$  with vector-like gauge couplings, as well as two  complex scalars, $\phi$ and $\chi$.  Interactions with the SM are mediated by a new gauge boson $Z'$ associated with an abelian U(1)$_X$ symmetry, under which only the particles of the dark sector are charged. The quantum numbers of all new particles are listed in TAB.~\ref{charges}~\footnote{A $Z'$ model that explains the $b\to s\mu^+\mu^-$ anomalies with tree-level couplings to quarks and leptons and provides an inherent DM candidate has recently been proposed in REF.~\cite{Sierra:2015fma}. Unlike in our model, the dark leptons mix with SM leptons, which results in a different phenomenology of dark matter and lepton observables. In particular, the $g_{\mu}-2$ contribution in the model in REF.~\cite{Sierra:2015fma} is typically much smaller than the needed shift in Eq.~\eqref{gm2data}.}.
 Since no chiral fermion carries U(1)$_X$
 charge, the model is 
 anomaly free~\cite{effectiveZprime}. 
Notice that our choice of U(1)$_X$ charges forbids tree-level $Z'$ couplings with SM leptons.
  This is  crucial
 in order to simultaneously address the 
  $g_{\mu}-2$ and $b\to s\mu^+\mu^-$ anomalies with NP around the weak scale.

Besides canonical kinetic terms, the relevant new interactions in the Lagrangian are 
\beq\label{NPops}
\mathcal{L}_{\rm NP}&\supset& \epsilon B_{\mu\nu}X^{\mu\nu}-\lambda_{\chi H}|\chi|^2|H|^2-\lambda_{\phi H}|\phi|^2|H|^2\nn\\
&&-V(\phi,\chi)-\left[y\,(\bar l   L)\chi + w\,(\bar q  Q)\phi +\hc\right]\,, 
\eeq
where
\beq
V(\phi,\chi)\equiv( r\phi\chi^2 +\hc)+ \lambda_\phi |\phi|^4+\lambda_\chi |\chi|^4\,, 
\eeq
and $l^T=\left(\nu_{\ell L}\,,\ell_L\right)^T$, $q^T=\left(u_L,d_L\right)^T$ and $H$ are the SM lepton, quark and Higgs doublets, while $B_{\mu\nu}$ and $X_{\mu\nu}$ are the hypercharge and U(1)$_X$ field strength tensors, respectively.  
The U(1)$_X$ symmetry is spontaneously broken by the vacuum expectation value (vev) of $\phi$,
 leading to a $Z'$ mass of $m_{Z'} = 2g'\langle \phi\rangle$. It furthermore lifts the mass degeneracy between the components of $\chi=(\chi_0+i\chi')/\sqrt{2}$ by
\beq\label{eq:scalar-splitting}
\delta \equiv \frac{m_{\chi'}^2}{ m_{\chi_0}^2}-1 = -2\frac{r\langle \phi\rangle}{m_{\chi_0}^2}.
\eeq
 We further assume that the singlet $\chi$
 is {\it inert}, {\it i.e.} does not develop a vev.
 There thus remains an exact $\mathbb{Z}_2$ symmetry,
 under which only $\chi$ and $L$ are odd,  which ensures that the lightest state of the spectrum is stable.
 For definiteness, we choose $r<0$ and $m_L > m_{\chi_0}$, so that
 the scalar component of $\chi$, $\chi_0$, is a DM candidate~\footnote{Choosing the neutral dark lepton $L^0$ as the lightest state of the spectrum would {\it a priori} also yield a DM candidate. However, it would already have been observed in direct DM detection experiments through $Z-$boson mediated interactions with nuclei.}.
 As we will argue in the
following sections, the above framework accommodates the $g_\mu-2$ and
 $b\to s\mu^+\mu^-$
 anomalies simultaneously, without conflicting with current collider constraints, only in particular
regions of parameter space, where the DM candidate $\chi_0$ is 
 around the weak scale and
 relatively lighter
 than the other dark sector states,
\beq
m_{\chi_0}
\lesssim  m_{\chi'}\sim m_{Q}\sim m_{Z'}\,,\label{spectrum}
\eeq
while the dark lepton  mass $m_L$ may be either close
 to the DM state or well above it.
\begin{table}[!t]
\caption{New fields and their quantum numbers. All SM fields are neutral under U(1)$_X$.}\vspace*{0.2cm}
\begin{tabular}{c|c|cccc}
 & spin & SU(3)$_c$ & SU(2)$_L$ & U(1)$_Y$ &  U(1)$_X$ \\
\hline \hline
$L$, $L^c$ & $1/2$& ${\bf1}$ &${\bf 2}$ & $-1/2$ & $1$\rule{0pt}{2.6ex}\\
$Q$, $Q^c$ & $1/2 $ & {\bf 3} & ${\bf 2}$ & $1/6 $ & $-2$\\
$\phi $ & 0&$ {\bf 1}$&${\bf 1}$ & $0$ & $2$\\
$\chi $ & 0&$ {\bf 1}$ & ${\bf 1}$
 & $0$ & $-1$  \rule[-1.2ex]{0pt}{0pt}\\
\hline
\end{tabular}
\label{charges}
\end{table}
 We do not address the dynamical origin of such a spectrum, but simply achieve it
by tuning the bare mass squared of $\chi$ 
 and $r\langle \phi\rangle$. We use $1/\delta$ as a measure of this tuning, which
 remains mild in the parameter region of interest.
 Besides addressing the anomalies,
 this small hierarchy of the spectrum also ensures that the lightest dark-sector state is a spinless SM-singlet DM candidate.

We now turn to the new interactions in Eq.~\eqref{NPops}. The first term
 describes kinetic mixing between the hypercharge U(1)$_Y$ and the dark U(1)$_{X}$ field tensors, which is strongly constrained by electroweak precision measurements~\cite{Kumar:2006gm} and collider searches~\cite{Chun:2010ve}. The second term, the scalar Higgs portal, is bounded by direct detection and collider searches~\cite{Djouadi:2011aa}. Since both interactions are not relevant to our discussion, we set $\epsilon=\lambda_{\chi H}=0$~\footnote{Even if absent at the tree-level, kinetic mixing is loop-induced through $B_{\mu}-X_{\mu}$ vacuum polarization effects with exchanged dark fermions $L$ and $Q$. While for the charges assumed in TAB.~\ref{charges}, these loop contributions are logarithmically UV-sensitive~\cite{HoldomU1}, they can be made finite in complete models by introducing additional vector-like fermions with appropriate U(1)$_X$ and SM charges~\cite{effectiveZprime}.}.

\begin{figure}[t]
\includegraphics[width=0.25\textwidth]{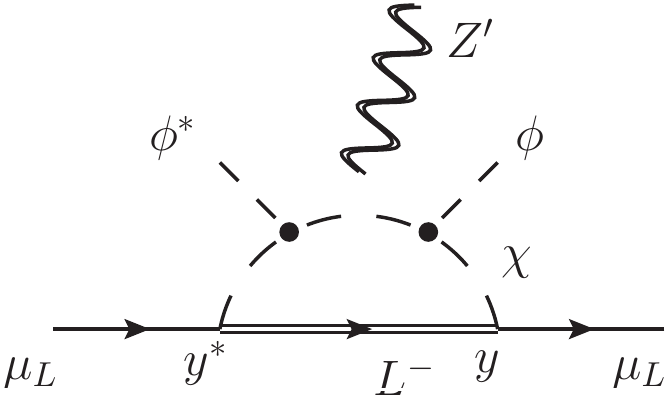}
\caption{Dominant one-loop contribution to the operator $(\phi^* D_\rho \phi)(\bar \mu_L \gamma^\rho \mu_L)$ yielding $\Gamma_{\mu\mu}(0)\neq 0$.
 The $Z'$ line is understood to be attached wherever possible. A similar diagram with $\mu\to \nu_{\mu}$ and $L^-\to L^0$ induces the $Z'$
 coupling to neutrinos.
}
\label{Zpmumu}
\end{figure}

 The last two terms
 mediate interactions between the dark sector and the SM.
The $Z'$  couples with strength $g'$ to the current $j^\rho\equiv j^\rho_0 +\delta j^\rho$, where
$j_0^\rho$ is the tree-level U(1)$_X$ current and
\beq  
\delta j^\rho \equiv\sum_{f,f'}\frac{\Gamma_{ff'}(q^2)}{1+\delta_{ff'}}\left(\bar f_L\gamma^\rho f'_L\right) + \hc \label{eq:u1x-current}
\eeq
is the part of the current induced after U(1)$_X$ breaking.  Here $q^2$ is the squared $Z'$ momentum, $f,f'=\ell,\nu_{\ell},u,d$, and $\delta_{ff'}=1$ for $f=f'$ and zero otherwise. SM quarks mix with dark quarks $Q$ after U(1)$_X$ breaking, while SM leptons and dark leptons $L$ do not mix due to the different charge assignments under U(1)$_X$ (see TAB.~\ref{charges}).
 The $Z'$ coupling to SM quarks  thus
 arises at tree level through the mixing
 $w(\bar q Q) \langle \phi \rangle +\hc$.
It is $q^2$-independent and isospin universal at leading order. In contrast, the $Z'$ coupling to leptons is
 radiatively induced at the one-loop level through the exchange of dark-sector fields via the interaction
 $y(\bar l L)\chi+\hc$, see FIG.~\ref{Zpmumu}.
 In the limit $m_{\ell,\nu_{\ell}}\ll m_{\chi,L}$, which we envisage here, the lepton form factors satisfy $\Gamma_{\ell\ell}(q^2)\simeq \Gamma_{\nu_\ell \nu_\ell}(q^2)$, since the isospin components of $L= (L^0,L^-)$ are
mass-degenerate at leading order. Calculating the one-loop diagrams of FIG.~\ref{Zpmumu} in the limit $q^2=0$ yields
\beq\label{Zpmu}
\Gamma_{\ell\ell}(0)=\frac{|y|^2}{32\pi^2}F_{Z'}(\tau, \delta)\,, 
\eeq
where $\tau\equiv m_L^2/ m_{\chi_0}^2$
 and the loop function $F_{Z'}$ is given in Eq.~\eqref{FZp}.

Notice that $F_{Z'}(\tau, \delta)$ vanishes in the limit of unbroken U(1)$_X$, $\delta\to 0$, where
 scalar and pseudo-scalar components  are degenerate. This stems from the fact that the form factor $\Gamma_{\ell\ell}(0)$ formally arises from the local operator $(\phi^*D_\rho\phi) (\bar \ell_L\gamma^\rho \ell_L)+\hc$ after U(1)$_X$ breaking,  induced by the one-loop diagram in FIG.~\ref{Zpmumu} with two insertions of the vev $\langle \phi \rangle$.

 As of the flavor structure of U(1)$_X$ interactions, we
 follow
 a phenomenological approach and introduce only interactions that are required to explain the observed
anomalies. We thus assume that the dark sector only couples to the left-handed fermions $\mu_L$, $\nu_{\mu L}$ and $\bar b_L s_L+\bar s_L b_L$.  The magnitude of the $Z' \bar{b}_L s_L$ coupling
 is experimentally constrained from
 $B_s$ meson mixing. Allowing for a NP contribution of $\mathcal{O}(10\%)$
 to the mass difference $\Delta M_{B_s}$
 leads to the bound~\cite{Buras-DeFazio-Girrbach,Altmannshofer-Straub}\footnote{This bound could be significantly weakened, if the $Z'$ also coupled to flavor-changing right-handed down-quark currents with a strength of $\mathcal{O}(10\%)$ of their left-handed counterparts~\cite{Buras-DeFazio-Girrbach}. We do not consider such a scenario here.}
\beq\label{DeltaMB}
|\Gamma_{bs}|\lesssim 2.4\times 10^{-3} \left(\frac{m_{Z'}/g'}{300\,\rm GeV}\right)\,.
\eeq
 We do not address here the origin of these peculiar flavor structures in the lepton and quark sectors. We acknowledge, however, that explicit flavor completions of this model could
 lead to correlated effects in other meson-physics and leptonic observables~\cite{Altmannshofer:2014cfa,Glashow:2014iga,Crivellin1,Crivellin2,MLFVbs}.

\section{Explaining Muon-related Anomalies}\label{sec:anomalies}
\begin{figure}[!t]
\includegraphics[width=0.25\textwidth]{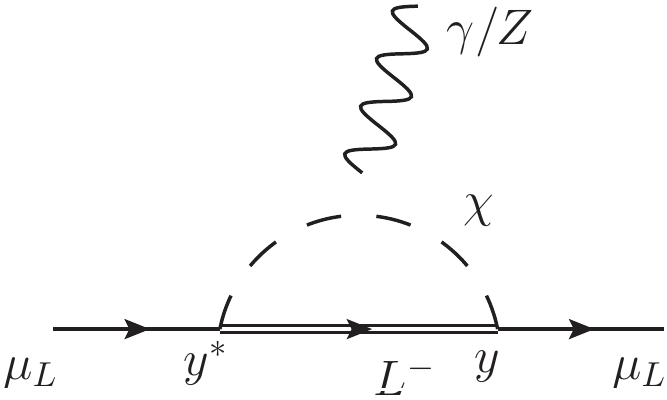}
\caption{One-loop NP contribution to $g_\mu-2$ and the $Z$ coupling to muons. The photon $\gamma$ is to be attached to $L^-$, and the $Z$ boson to either of the fermion lines. 
}
\label{gm2diagram}
\end{figure}

New contributions to the muon anomalous magnetic moment
are induced at the one-loop level through the diagram
in FIG.~\ref{gm2diagram}, yielding
\beq\label{gm2}
\Delta a_\mu^{\rm NP} &=& \frac{|y|^2}{32\pi^2}\frac{m_\mu^2}{m_{\chi_0}^2}F_{g}(\tau)\left[1+A(\tau,\delta)\right]\,,
\eeq
with $m_\mu$ the muon mass and
$
 A(\tau,\delta)\equiv \frac{F_{g}(\tau/(1+\delta))}{(1+\delta) F_{g}(\tau)}$. The loop function $F_g$ is given in Eq.~\eqref{Fg}.   For a
 degenerate dark spectrum, we have $F_{g}(1)=1/12$ and $A(1,0)=1$, so that the discrepancy $\Delta a_{\mu}$ in Eq.~\eqref{gm2data} is accommodated for $m_{\chi_0}\simeq45|y|\,$GeV.
 As we will see, the LHCb anomalies typically require a significant scalar versus pseudo-scalar mass splitting
$\delta \gg1$, with a dark-lepton versus DM mass splitting roughly in
 the interval $\tau\in [1,\delta]$.
 Eq.~\eqref{gm2data} is then accommodated for
 $m_{\chi_0}\simeq 32|y|(1+2/\delta)\,$GeV ($\tau\simeq 1$) or $m_{\chi_0}\simeq 55|y|/\sqrt{\delta}\,$GeV  ($\tau\simeq \delta\gg 1$). (See APP.~\ref{App} for details.)

\begin{figure}[!t]
\includegraphics[width=0.35\textwidth]{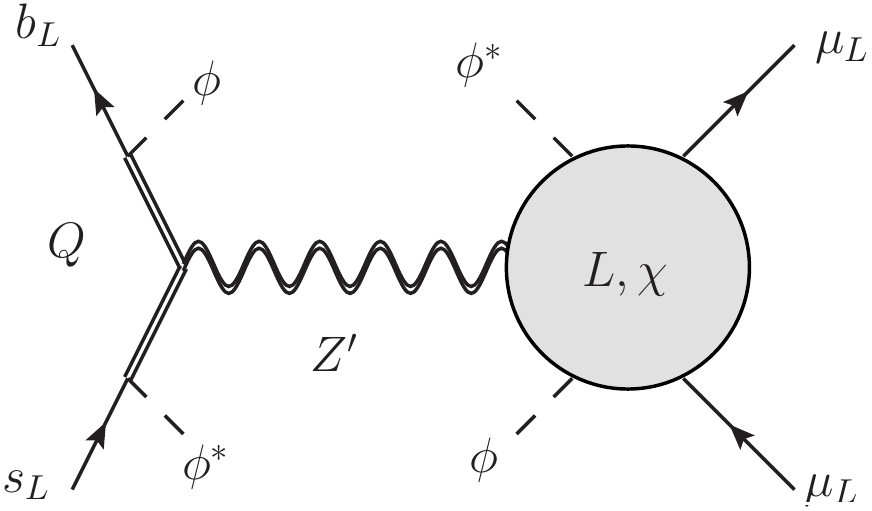}
\caption{Leading NP contribution to the Wilson coefficients $C_{9,10}^\mu$. The shaded disk denotes the radiatively induced $Z'$ coupling to muon pairs
 at zero momentum exchange, as shown in FIG.~\ref{Zpmumu}.}
\label{darkZ_bsmumu}
\end{figure}

Consider now the $b\to s\mu^+\mu^-$ anomalies.
At the $b-$quark mass scale, the NP amplitude
 is described by the effective Hamiltonian
\beq
\mathcal{H}^{\rm NP}_{\rm eff} = -\frac{\alpha G_F}{2\sqrt{2}\pi}V_{tb}V_{ts}^*\sum_i C_i \mathcal{O}_i+\hc\,,
\eeq 
where $\alpha$ 
and $G_F$ are, respectively, the fine-structure and Fermi constants, $V_{ij}$ are CKM matrix elements, and the sum runs over the operators ($\ell = e,\mu$)
\beq
\mathcal{O}_9^\ell &\equiv& \bar b \gamma_\rho(1-\gamma_5)s\ \bar \ell \gamma^\rho \ell\,,\\
\mathcal{O}_{10}^\ell &\equiv& \bar b \gamma_\rho(1-\gamma_5)s\ \bar \ell \gamma^\rho\gamma_5 \ell\,.
\eeq
 A global fit to leptonic, semi-leptonic and radiative $B$ decays favors effective couplings to muons~\cite{Hiller:2014yaa,Altmannshofer:2015sma}
\beq\label{eq:fit}
C_9^\mu = -C_{10}^\mu \simeq-0.5
\eeq
and negligible electron
coefficients $C_{9,10}^e\simeq 0$~\footnote{A slightly better fit is obtained for a pure vector 
 NP coupling to muon pairs, {\it i.e.} for $C_9^\mu\simeq-1$ and $C_{10}^\mu\simeq 0$~\cite{Altmannshofer:2015sma}. In scenarios, where $C_9^{\mu}$ is induced at the loop level (as in our model), it is difficult to achieve such a large effect. However, from a model-building perspective the solution $C_9^{\mu}=-C_{10}^{\mu}\approx-0.5$ with chiral interactions is more inviting, as it allows a natural implementation of weak isospin gauge invariance.}.
 In
 our model, $C_{9,10}^{\mu}$ are induced at the one-loop level through the diagram in FIG.~\ref{darkZ_bsmumu}, which gives
\beq\label{C910}
C_{9}^{\mu} = -C_{10}^{\mu} = - \frac{g'^2}{4m_{Z'}^2}\Lambda_{\rm SM}^2
 \frac{|V_{tb}V_{ts}^*|}{V_{tb}V_{ts}^*}\, \Gamma_{bs}\Gamma_{\mu\mu}\,,
\eeq
where $\Lambda_{\rm SM}=[2\sqrt{2}\pi/(\alpha G_F |V_{tb}V_{ts}^*|)]^{1/2}\simeq 50\,$TeV
 is the scale of the SM contributions. Accommodating the $b\to s\mu^+\mu^-$ anomalies as in Eq.~\eqref{eq:fit}, while respecting the
 $\Delta M_{B_s}$ bound from Eq.~\eqref{DeltaMB}, yields a lower bound on the muon form factor,
\beq\label{xmu}
\Gamma_{\mu\mu}(m_b^2)\gtrsim 0.029\left(\frac{m_{Z'}/g'}{300\,{\rm GeV}}\right)\,.
\eeq
Since the $Z'$ coupling to muons is small, {\it cf.} Eq.~\eqref{Zpmu}, this bound is fulfilled only for a light $Z'$ boson, as well as a large mass splitting $\delta$. 
 In a spectrum with moderate tuning, $\delta\lesssim 10$, this leads to
\beq\label{Zpbound}
m_{Z'}\lesssim 110-270\,{\rm GeV}\left(\frac{g'}{3}\right)\left(\frac{|y|}{3}\right)^2\,,
\eeq
where the two mass values
 correspond to $\tau\simeq \delta$ and $\tau\sim1$, respectively. For $m_{Z'}$ around the weak scale,
 LHCb anomalies thus
 require rather large couplings
 $y$ and $g'$,
 while a perturbative upper
 bound of $m_{Z'}\lesssim 8.5\,$TeV  ($\tau\simeq \delta$) or $m_{Z'}\lesssim 20\,$TeV  ($\tau\simeq 1$) applies for $g'=y=4\pi$. We discuss collider constraints on such a $Z'$ in SEC.~\ref{collconst}.
 To summarize, assuming perturbative couplings $g'\simeq 3$ and a mild tuning of the dark spectrum, $\delta \simeq 10$, both the $g_{\mu}-2$ and LHCb anomalies are accommodated for parameters values interpolating between the two limiting cases
\beq\label{sweetspot}
1)&&\tau\simeq\delta\,,\ |y|\simeq 6\,,\  g'\simeq 3\,,\ \delta \simeq 10\,,\nonumber\\
&&m_{\chi_0}\simeq 100\,{\rm GeV},\ m_{Z'}\simeq 300\,{\rm GeV},\
\eeq
\beq\label{sweetspot2}
2)&& \tau\simeq 1\,,\ |y|\simeq 2\,,\ g'\simeq 3\,,\ \delta \simeq 10\,,\nonumber\\
&& m_{\chi_0}\simeq 70\,{\rm GeV},\ m_{Z'}\simeq 150\,{\rm GeV}.\ \,
\eeq

\section{Collider Constraints}\label{collconst}
We now analyse
 the relevant constraints on our model from EW precision measurements at the LEP experiments, as well as from the first LHC run.
First of all, our model implies sizeable radiative corrections to the $Z\mu\bar \mu$, $Z\nu_\mu\bar \nu_\mu$ and $W^+ \mu\bar \nu_{\mu}$ couplings  
from the one-loop diagram in FIG.~\ref{gm2diagram}. The EW gauge couplings are shifted by ($V=W,Z$)
\beq
\frac{\delta g}{g_{\rm SM}}= \frac{|y|^2}{32\pi^2}F_V(\tau,r_q)\,,
\eeq
where the one-loop function $F_V$ is found in Eq.~\eqref{FV} and $r_q\equiv q^2/m_L^2$. The vertex correction is $m_V^2/m_L^2$-suppressed at the $V$ pole, $q^2=m_V^2$. While the QED part of the
 couplings at zero momentum is protected by gauge invariance, the weak isospin part is not corrected at one-loop level, since the SM Higgs doublet does not directly couple to the dark sector~\footnote{This implies that the operator $(H^\dagger D_\rho H)\bar l \gamma^\rho l+\hc$ (relevant for the $Z\mu^+\mu^-$ coupling), as well as  $H^\dagger \sigma\cdot W_{\mu\nu} H B^{\mu\nu}$ and $|H^\dagger D_\mu H|^2$ (which respectively shift the $S$ and $T$ parameters~\cite{PeskinSTU}), are only induced at the two-loop level. Effects on $S$ and $T$ are thus sufficiently small to evade the LEP bounds.}. Despite this parametric suppression, the lightness of the dark-sector states, $m_{\chi_0}\simeq 100\,$GeV and $m_L\simeq 100-400\,$GeV, together with a relatively large Yukawa coupling, $y\gtrsim 2$, typically shifts the SM gauge couplings
 by one to a few permil, which is in mild tension ($\sim1\sigma-3\sigma$) with LEP data~\cite{PDG}. This tension
may be relieved
 in a more sophisticated version of the minimal model considered here. 

Our model also predicts a series  of signatures at hadron colliders, most notably muon pair production through a resonant 
 $Z'$, as well as signals of large missing energy
 with muon pairs and/or jets. 
Current LHC limits on a $Z'$ resonance with SM-like couplings to fermions are around $m_{Z'}\gtrsim3\,$TeV~\cite{Aad:2014cka,Khachatryan:2014fba}. However, $Z'$ production in our model only occurs through sea-quark ($b\bar s+s\bar b$) annihilations and is thus strongly suppressed. On the other hand,    
 the $Z'$ dominantly decays into muon pairs and neutrinos. Given the conditions Eqs.~\eqref{DeltaMB} and~\eqref{xmu} on the $Z'$ couplings to SM fermions and for $Z'$ masses that accommodate the LHCb anomalies as in Eq.~\eqref{Zpbound}, the branching ratios to leptons are both $\sim 40-50\%$, depending on the value of the Yukawa coupling~\footnote{We use CalcHEP 3.4~\cite{CalcHEP34} to compute the $Z'$ production cross sections and partial decay widths assuming the spectrum in Eq.~\eqref{sweetspot}.}. We thus find the cross section for $Z'$ production  with decay into
 $\mu^+\mu^-$ to be of $\mathcal{O}$(fb), which is an order of magnitude below current limits at the $8\,$TeV LHC. The next-to-leading $Z'$ branching ratio
 is $\mu^+\mu^-\chi_0\chi_0$, ranging from 2 to 10$\%$
 for large Yukawa couplings.  Along the same lines, mono-jet signatures from
 the direct production of a DM pair in association
 with  a hard jet from initial state radiation (ISR)
 lie at least one order of magnitude below the current LHC sensitivity~\cite{Aad:2015zva,Khachatryan:2014rra}.

The EW production of $L^+L^-$ pairs leads to a signature with di-muons and missing energy, which resembles the one used in searches for
smuons, the supersymmetric partners of the muon. The only difference with our signal lies in the spin of the produced particles. However, it was shown in REF.~\cite{Arina:2015uea} that
 results for slepton searches in simplified models could 
 safely be applied to the production of fermion pairs decaying into a fermion and a scalar. We use {\tt SmodelS}~\cite{Kraml:2014sna}, a tool designed to decompose the signal of any NP
 model into simplified topologies, and compare the predictions to the exclusion limits set by the ATLAS and CMS slepton searches~\cite{Aad:2014vma,Khachatryan:2014qwa}.
 We find that the 8$\,$TeV LHC sets
 strong constraints on the mass of $L^-$, even stronger than for smuons,
 because $L^+L^-$ pair production cross sections are significantly larger. Dark-lepton masses $m_L\lesssim450\,$GeV are excluded, except if the mass splitting with the DM is sufficiently small, $m_L-m_{\chi_0}\lesssim60\,$GeV. In this region, the di-muon signal is overwhelmed with SM background.  Similar searches at LEP2 lead to the lower bound of $m_L\gtrsim 100\,$GeV~\cite{Abdallah:2003xe}.

\section{Dark Matter Relic Abundance}\label{DM}
\begin{figure}[!t]
\begin{tabular}{ccc}
\includegraphics[width=0.20\textwidth]{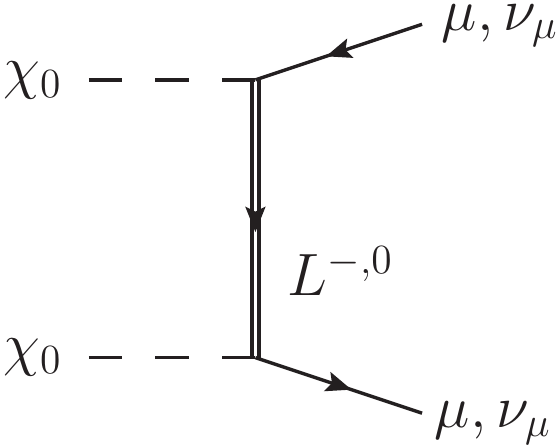}&\hspace{0.7cm} &\includegraphics[width=0.20\textwidth]{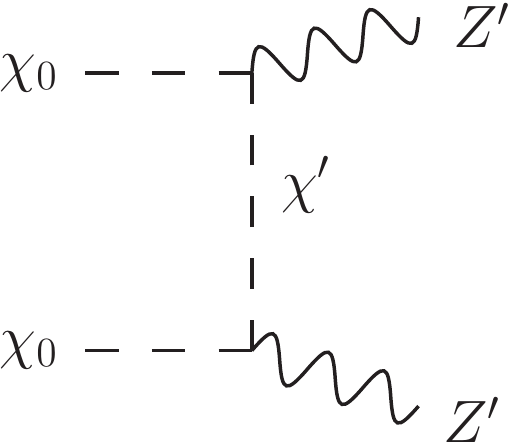}\\
&&\\
\includegraphics[width=0.20\textwidth]{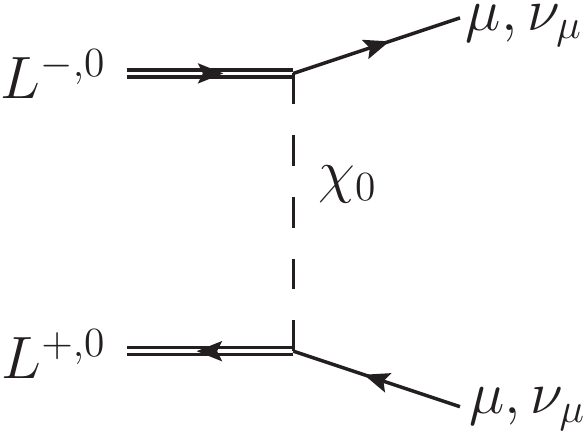}&&\includegraphics[width=0.20\textwidth]{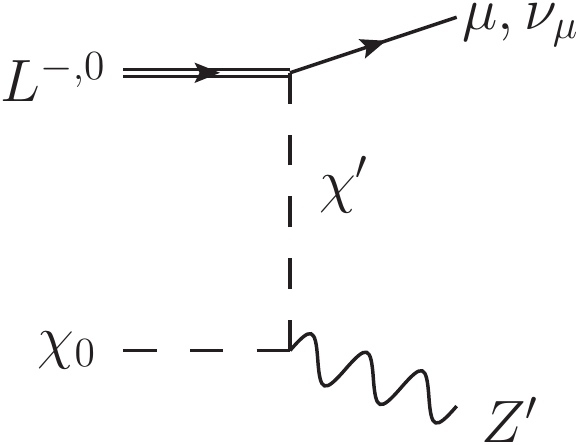}
\end{tabular}
\caption{Dominant amplitudes for DM-DM annihilation (top) and DM-$L$ or $L$-$L$ coannihilation (bottom) in the early universe.}
\label{DMdiagrams}
\end{figure}

The DM candidate in our model is the lightest component of the scalar $\chi$, which we assume to be $\chi_0$. It is largely leptophilic, as follows from the charge assignments in TAB.~\ref{charges}.
For a spectrum as in
 Eq.~\eqref{spectrum}, DM annihilation proceeds dominantly
 into $\mu^+\mu^-$ and $\nu_\mu\bar\nu_\mu$
 through $t-$channel exchange of $L^-$ and $L^0$, respectively, as shown in FIG.~\ref{DMdiagrams}.
Coannihilation processes with the other dark states are negligible for $\tau\simeq\delta\gg1$, while annihilation into $Z'$ pairs is negligible as long as $m_{Z'}\gtrsim m_{\chi_0}$.
 The resulting annihilation cross section is $d-$wave suppressed in the chiral limit,
 $\sigma_{ \bar l  l}v \sim v^4$~\cite{ScalarDMTytgat}. Adding up final-state muons and neutrinos, the thermal average is~\footnote{In our analytic discussion, we work 
 at leading order in $v\simeq 0.1-0.3$ at freeze-out in a non-relativistic expansion, while in our numerical analysis the relic density is computed with {\tt micrOMEGAs}~\cite{MO41}.}
\beq\label{2body}
\langle \sigma_{\bar l l} v\rangle = \frac{a_d}{x^2 m_{\chi_0}^2} \,+\, \mathcal{O}(x^{-3})\,,\quad a_d\equiv \frac{|y|^4}{2\pi(1+\tau)^4} \,, 
\eeq
where $v$ is the relative DM-DM velocity, and $x\equiv m_{\chi_0}/T$, the DM mass-to-temperature ratio. For the parameter set in Eq.~\eqref{sweetspot}, $a_d\simeq 0.02$ is a typical value. In the freeze-out approximation~\cite{GelminiGondolo}, the relic density is 
\beq\label{Oh2}
\Omega_\chi^{ \bar l l} h^2\simeq 8.5\times 10^{-11}\,\left(\frac{3x_f^3 m_{\chi_0}^2}{a_d\sqrt{g_*}{\rm GeV}^2}\right)\times[1+R_f]^{-1},
\eeq
where $g_*\simeq80$ is the effective number of relativistic degrees of freedom at $T_f
\simeq 1-10$~GeV and 
\beq\label{Yfcorr}
R_f\simeq 2.6\times 10^{-18}\left(\frac{\sqrt{g_*}m_{\chi_0}}{a_d{\rm GeV}}\right)x_f^{3/2}\exp(x_f)
\eeq
accounts for the reduced DM abundance at freeze-out.
 Since DM annihilation
 $\langle \sigma_{\bar l l} v\rangle\propto x^{-2}$ is $d-$wave suppressed during freeze-out, DM decoupling is slightly delayed, which results in a larger (power of) $x_{f}\simeq 25$ in Eq.~\eqref{Yfcorr} compared to $s-$ or $p-$wave annihilation. 
 In a scenario as in
 Eq.~\eqref{sweetspot}, $R_f\simeq \mathcal{O}(1)$ is a significant correction.

Without this correction and for fixed mass splittings $\tau$ and $\delta$, the relic density 
 scales as $\Omega_{\chi}^{\bar l l}h^2\propto m_{\chi_0}^2/|y|^4$, {\it i.e.} like $(\Delta a_\mu \times C_9^\mu)^{-1}$, in our model. Using Eqs.~\eqref{gm2} and~\eqref{C910}, Eq.~\eqref{Oh2} reads ($R_f=0$)
\beq\label{relic-prediction}
\Omega_\chi^{\bar l l} h^2&\simeq&0.01\times I(\tau,\delta)\left(\frac{x_{f}}{25}\right)^3\left(\frac{100\,{\rm GeV}}{m_{Z'}/g'}\right) \nn\\
&&\times\left(\frac{287\times 10^{-11}}{\Delta a_\mu}\right)\left(\frac{0.5}{|C_9|}\right),
\eeq
where $I(\delta,\tau)\equiv F_g(\tau)\left[1+A(\tau,\delta)\right]F_{Z'}(\tau,\delta)(1+\tau)^4$. Assuming parameter values of Eq.~\eqref{sweetspot} (with $\tau\gg 1$) in order to accommodate the muon anomalies yields $I(\tau,\delta)\simeq 20-40$. Recalling that $R_f\sim \mathcal{O}(1)$, this gives
 a relic abundance close to the observed value from Planck data~\cite{Ade:2015xua} 
\beq
\Omega_\chi^{\rm obs} h^2 = 0.1199\pm 0.0022\,.
\eeq 
However, for parameter values of Eq.~\eqref{sweetspot2} (with $\tau\simeq 1$), $I(\tau,\delta)\simeq \mathcal{O}(1)$ and the resulting relic density is typically a factor of $5-10$ smaller than observed.
We stress that the $d-$wave suppression of scalar DM annihilation into lepton pairs through a heavy fermion mediator plays an important role in the above prediction. Had DM annihilation been $p-$wave ($s-$wave) dominated, the same parameter space would have predicted a relic density smaller by a factor of about ten (hundred).

Since the two-body process in Eq.~\eqref{2body} is strongly suppressed~\footnote{Contributions from the finite muon mass, which induce $s-$wave annihilation into di-leptons, are of $\mathcal{O}(10^{-3})$ and thus negligible.}, parametrically sub-dominant annihilation channels can
 become relevant. The leading such contribution is three-body annihilation $\chi_0\chi_0\to \mu^+\mu^-\gamma$ with an extra photon in the final state. Virtual internal Bremsstrahlung (VIB) from the dark lepton $L^+$
 lifts the kinematic suppression, so that annihilation proceeds through an $s-$wave~\cite{Bringmann:2007nk,ScalarDMTytgat}. The thermally averaged cross section for this process is~\cite{Toma:2013bka}
\beq
\langle \sigma_{\mu\bar\mu\gamma}v\rangle = \frac{\alpha}{32
\pi^2}\frac{|y|^4}{m_{\chi_0}^2}F_{\gamma}(\tau)\,,
\eeq
with the function $F_{\gamma}$ defined in Eq.~\eqref{3bodycorr}.
Following REF.~\cite{Toma:2013bka}, we find that VIB
 suppresses the predicted relic density in Eq.~\eqref{Oh2} by 
 a factor 
 $R_{\gamma}\equiv \Omega_\chi^{\bar ll+ \mu\bar\mu\gamma}/\Omega_\chi^{\bar l l}<1$,
 which in the chiral limit and at zero relative DM velocity $v=0$ approximates~\footnote{This expression is only an estimate of the total $\mathcal{O}(\alpha)$ correction to DM
 annihilation into the
 final state $\mu^+\mu^-$. A complete calculation that takes account of
 the full DM-velocity dependence, requires to also include virtual corrections.
  We leave a thorough treatment of the extra-photon contribution for future implementation in
 {\tt micrOMEGAs}.}
\beq
R_\gamma^{-1} \simeq 1+\frac{3\alpha}{16\pi}\, x_f^2(1+\tau)^4F_\gamma(\tau)\,.
\eeq
The function $(1+\tau)^4F_\gamma(\tau)$ strongly decreases with $\tau$, yielding to suppression factor of $R_\gamma\simeq 0.5\ (0.9)$ for $\tau\simeq1$ ($\tau\gg 1$). Therefore, VIB is a significant contribution only in particular regions of parameter space, where coannihilation in addition strongly
 depletes the relic density. 
 
\begin{figure}[!t]
\includegraphics[width=0.45\textwidth]{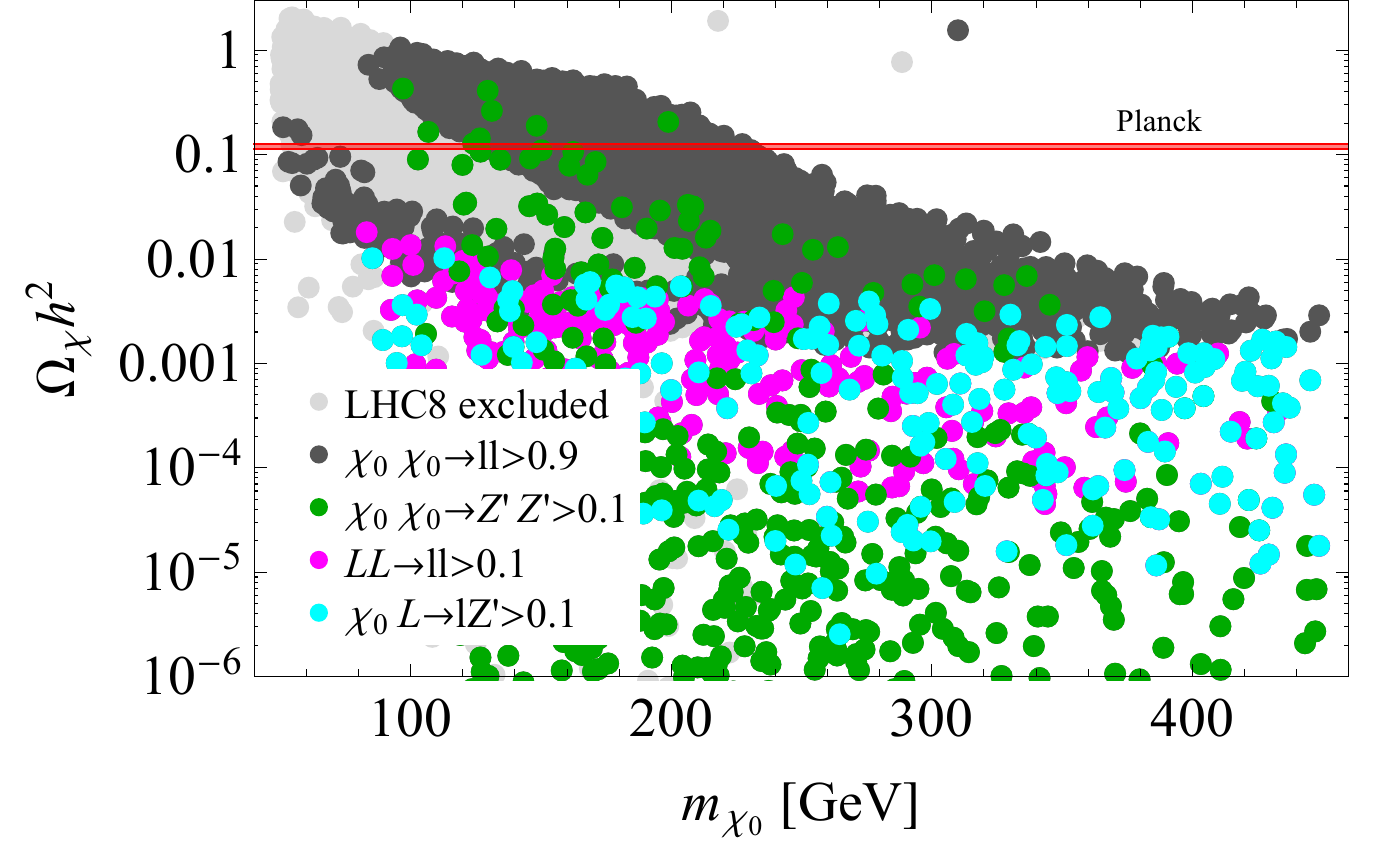}
\caption{DM relic density $\Omega_\chi^{\bar ll} h^2$ as a function of the DM mass, $m_{\chi_0}$, as predicted by the muon-related collider anomalies. Light gray points: excluded by 8TeV LHC and LEP2 constraints on di-lepton plus missing energy signals from $L^+L^-$ production
Dark gray points:
 DM annihilation
 dominated by $\chi_0\chi_0\to \mu^+\mu^-,\nu_\mu \bar{\nu}_\mu$. Green points: additional
 contributions ($\sim10\%$ and more) from $\chi_0\chi_0\to Z'Z'$. Magenta  and cyan points: co-annihilation through $LL\to ll$ and $\chi_0L\to Z'  l$, respectively, accounts for more than $\sim 10\%$ of the total $\langle \sigma v\rangle$. Red band: $3\sigma$ 
range of $\Omega_{\chi}^{\rm obs}$ from Planck~\cite{Ade:2015xua}.  (See text for details.)}
\label{relic}
\end{figure}

In order to assess the robustness of the
 prediction for the relic density in Eq.~\eqref{relic-prediction} more quantitatively, we broadly
 scan around the parameter sets in Eqs.~\eqref{sweetspot} and~\eqref{sweetspot2} that accommodate the muon anomalies, 
 varying the parameters in the following ranges
\beq\label{scant}   
50\leq m_{\chi_0}\leq450\,{\rm GeV}\,,\ 200\leq m_{Z'}\leq 1000\,{\rm GeV}\,,\nn\\
1\lesssim\tau\leq 20\,,\ 0\lesssim \delta\leq20 \,,
\ 1\leq|y|\leq4\pi \,,\ 1\leq g'\leq 5
\,.\ 
\eeq
We restrict the tuning amoung the scalar components to maximally $10\%$. Furthermore, we impose the condition that the $g_\mu-2$ and the LHCb anomalies quoted in SEC.~\ref{intro} are explained within one standard deviation, and that the collider limits on di-muon and missing energy production from SEC.~\ref{collconst} are satisfied. 
We then compute the resulting relic density with {\tt micrOMEGAs}~\cite{MO41}.
As shown in FIG.~\ref{relic},  for many parameter points where the dominant DM annihilation process is $\chi_0\chi_0\to \bar{l}l$ (dark gray points) the relic density falls
 within an order of magnitude of  $\Omega^{\rm obs}_\chi$, as expected from the above discussion. However, we also find that
 the predicted relic density can be much lower than
 the observed value, if the spectrum significantly deviates from Eq.~\eqref{spectrum}. First of all, 
 $\Omega_\chi$ rapidly decreases when the DM mass increases above $m_{\chi_0}\gtrsim 200\,$GeV.
 Satisfying $\Delta a_{\mu}$ with
 heavier DM masses requires a lighter dark lepton $L$ or, equivalently, smaller values of $\tau$, leading to a significantly stronger DM annihilation into muon(-neutrino) pairs, see Eq.~\eqref{2body}. Secondly,
 when $ m_{Z'} \lesssim m_{\chi_0} $, the process $\chi_0\chi_0\to Z'Z'$ efficiently depletes the DM relic abundance (green points in FIG.~\ref{relic}). 
Moreover, when $m_L\simeq m_{\chi_0}$,
 corresponding to $\tau\simeq 1$, coannihilation processes with dark leptons  (see FIG.~\ref{DMdiagrams}) can also contribute to significantly reducing the DM density (magenta
 points in FIG.~\ref{relic}). This is particularly pronounced when $Z'$ can be produced in the final state via the coannihilation processes $\chi_0L^\pm(L^0) \to Z' \mu^\pm (\nu_\mu)$ (cyan points in FIG.~\ref{relic}).

\begin{figure}[!t]
\includegraphics[width=0.45\textwidth]{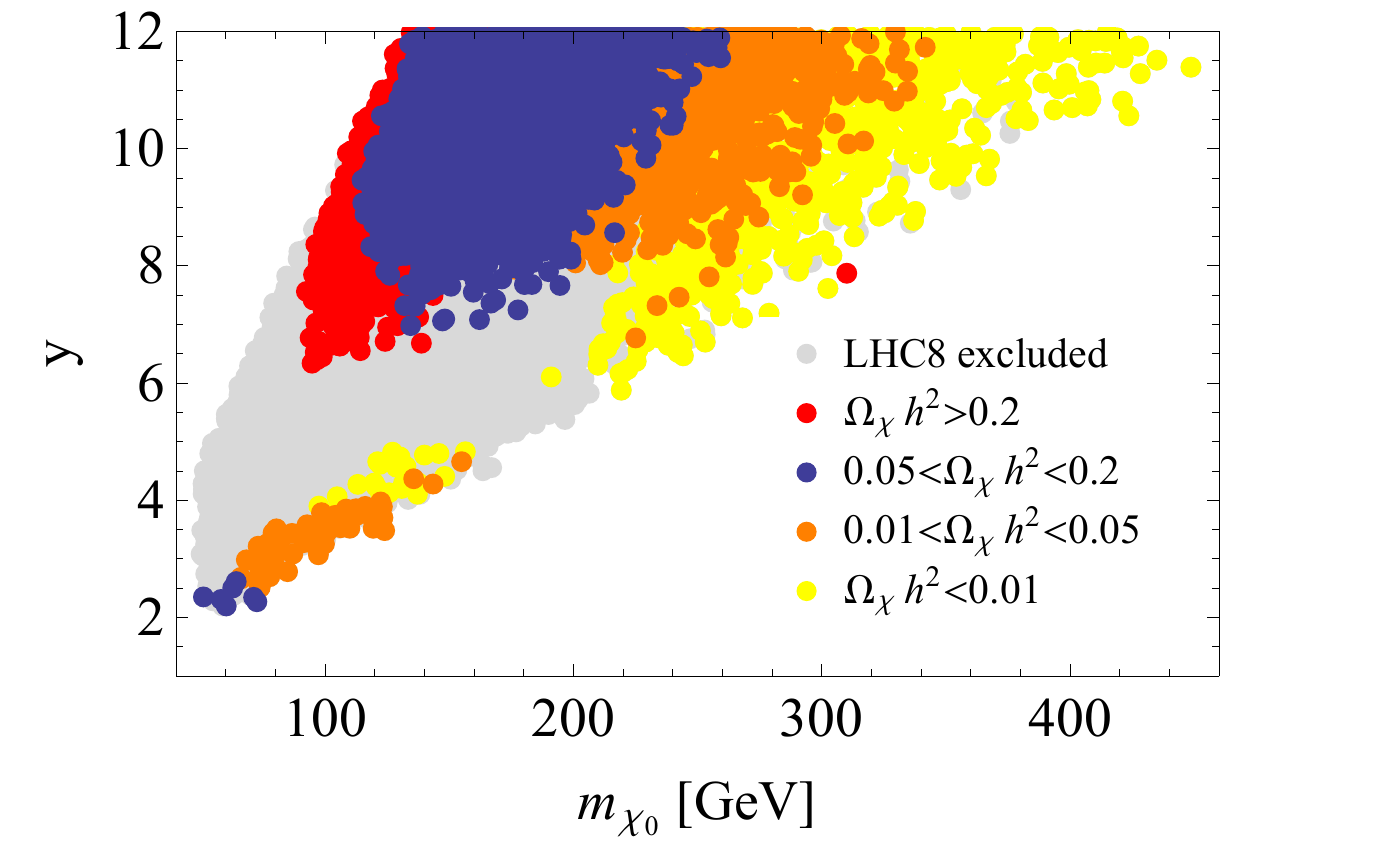}
\caption{Dark Yukawa coupling $y$ as a function of $m_{\chi_0}$ required to accommodate the muon
 anomalies within one standard deviation. Gray points are excluded by 8TeV LHC and LEP2 constraints on di-lepton plus missing energy signals
 from $L^+L^-$ production. All other points correspond to
 dominant DM annihilation $\chi_0\chi_0\to \mu^+\mu^-,\nu_{\mu}\bar{\nu}_{\mu}$, {\it i.e.} to the dark gray points in FIG.~\ref{relic}. Colors depict different amounts of the associated DM relic abundance; $\Omega_{\chi}^{\rm obs}h^2$ is obtained in the blue regions.}
\label{y-mX}
\end{figure}

The two regions where the relic density is in agreement with the observed value
 closely resemble the parameter limits favored by the muon anomalies from Eqs.~\eqref{sweetspot} and~\eqref{sweetspot2}, respectively.  These are 
\begin{itemize}
\item[1)] The {\it heavy lepton scenario}. In  this region,
 $m_{\chi_0}\approx 120-250\,$GeV and $m_L\gtrsim 450\,$GeV, corresponding to $\tau \gg 1$, because of collider constraints on di-lepton plus missing energy signals. As a result, the Yukawa coupling is strong, $|y|\gtrsim 6$, to ensure a large enough DM annihilation rate into lepton pairs.
\item[2)] The {\it compressed scenario}. This second region
 corresponds to $m_{\chi_0}\approx 50-80\,$GeV and $m_L\approx100-120\,$GeV, {\it i.e.} to $\tau\simeq 1$. It evades collider constraints because of the relatively small gap between DM and dark-lepton masses. 
 Since DM annihilation is larger for $\tau\simeq 1$, see Eq.~\eqref{2body}, a smaller
 Yukawa coupling is
 needed to ensure enough DM annihilation. However, as the muon anomalies request $|y|\gtrsim 2$, it is more challenging to
 obtain the observed relic density $\Omega_\chi^{\rm obs} h^2 \sim 0.1$ in the compressed scenario.
\end{itemize}
 In FIG.~\ref{y-mX}, we illustrate the range of $y$
 needed to obtain
 various amounts
 of DM relic abundance $\Omega_\chi$, as a function of the DM mass, for the 
 case of dominant DM annihilation into lepton pairs, $\chi_0\chi_0\to \bar l l$.

\section{Dark Matter Detection}\label{ID}
The DM candidate in our model is dominantly leptophilic, which makes direct detection very challenging. Furthermore, since $Z$ and Higgs boson couplings to $\chi_0$ pairs are
 absent at tree level and
 loop-induced vector interactions are momentum-suppressed, any signal of spin-independent DM-nucleus scattering lies below the sensitivity of current direct detection experiments~\footnote{The leading contribution to spin-independent (SI) scattering 
 arises at the two-loop level from the effective interaction $\chi_0\chi_0\bar q q$. For the spectrum in Eq.~\eqref{sweetspot}, we estimate its size to be $\sigma_{\rm SI}^p\sim \mathcal{O}(10^{-50})\,$cm$^2$, which lies below the coherent neutrino background~\cite{BillardCNB} for DM  masses around the weak scale. Spin-dependent interactions are strongly velocity-suppressed in the non-relativistic limit.}.

On the other hand, VIB  in DM annihilation, discussed in SEC.~\ref{DM}, plays an important role for indirect detection today. The suppression of tree-level DM annihilation into di-lepton states, $\sigma_{\bar ll} v \sim v^4$, is even stronger than during freeze-out, as velocities today are around $v_{\rm halo}\sim 10^{-3}$ in our galactic neighborhood. Photons from VIB are thus the dominant signal in gamma-ray searches of indirect detection experiments. The annihilation cross sections due to VIB in our two scenarios from Eqs.~\eqref{sweetspot} and \eqref{sweetspot2}, integrated over the photon energy spectrum, are
\beq\label{eq:indirect}
1) \quad\ \langle \sigma_{\mu\bar\mu\gamma}v\rangle && \hspace*{-0.25cm}\simeq 3\times 10^{-27}\,\text{cm}^3/\text{s}\quad \text{and}\nonumber\\
2) \quad\ \langle \sigma_{\mu\bar\mu\gamma}v\rangle && \hspace*{-0.25cm}\simeq 7\times 10^{-25}\,\text{cm}^3/\text{s}\,,
\eeq
respectively. The smaller cross section in the heavy lepton scenario 1) is mainly due to the suppression $\langle \sigma_{\mu\bar\mu\gamma}v\rangle\sim 1/\tau^4$ in the limit of large mass splitting, $\tau\gg1$.

For small mass splitting $\tau\simeq 1$, the energy spectrum of VIB photons is peaked close to the endpoint $E_{\gamma}=m_{\chi_0}$~\cite{Bringmann:2007nk}, which leaves a characteristic signature in indirect detection experiments. The Fermi satellite has performed a dedicated search for spectral features in VIB using data from space regions near the galactic center collected during Pass 7~\cite{Bringmann:2012vr}. Assuming a spectral shape characteristic for $1\lesssim \tau \lesssim 2$, this search sets an experimental upper bound on VIB,
\beq
\langle \sigma_{\mu\bar\mu\gamma}v\rangle \lesssim 10^{-27}\,\text{cm}^3/\text{s}\,,
\eeq
for $m_{\chi_0}\lesssim 100\,$GeV.

 More inclusive analyses of Fermi data from dwarf galaxies constrain the overall gamma-ray flux emitted in DM annihilations~\cite{Ackermann:2011wa,Ackermann:2015zua}. These searches are sensitive to the sum of secondary photons (from final-state radiation off muons and muon decays) and VIB. Though the cross section bounds are generally weaker than from the dedicated VIB search, they constrain $\langle \sigma_{\mu\bar\mu\gamma}v\rangle$ for scenarios with large $\tau$, which the VIB spectral search is not sensitive to.

\begin{figure}[!t]
\includegraphics[width=0.45\textwidth]{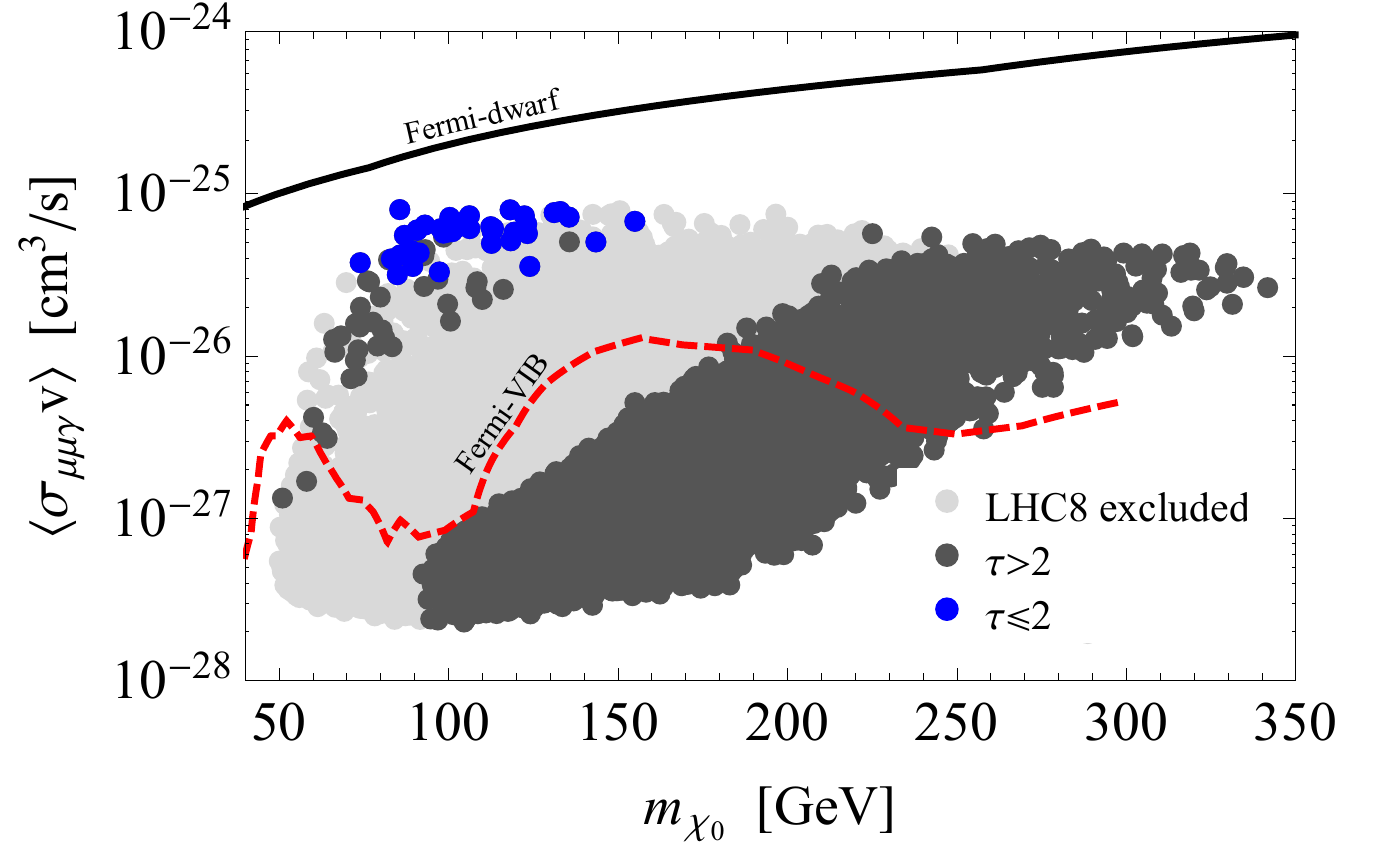}
\caption{Thermally averaged cross section for DM annihilation into $\mu^+\mu^-\gamma$ in our galactic neighborhood today as a function of the DM mass,  as predicted by muon-related collider anomalies. All points satisfy $0.01\lesssim \Omega_\chi h^2\lesssim 1$. Light gray points are excluded by 8TeV LHC and LEP2 constraints on di-lepton plus missing energy signals
 from $L^+L^-$ production. Points with $\tau\leq 2$ ($\tau > 2$) are shown in blue (dark gray). The black line denotes the Fermi LAT upper bound on photon emission in nearby dwarf galaxies, while the dashed red line represents the 
 Fermi LAT limit on spectral features in VIB signals~\cite{Bringmann:2012vr}, which applies only for $1 \lesssim \tau \lesssim 2$, \emph{i.e.}, to the blue points.}
\label{vib-mx}
\end{figure}

In FIG.~\ref{vib-mx}, we compare the predictions of DM annihilation through VIB in our model with the bounds from Fermi.
We focus only on points in the scan range of Eq.~\eqref{scant}, which satisfy the muon-related anomalies and yield a DM relic density of $0.01 < \Omega_{\chi}h^2 < 1$.
 The upper bound on $\langle \sigma_{\mu\bar\mu\gamma}v\rangle$ from Fermi's VIB search was derived in REF.~\cite{Bringmann:2012vr} (dashed red line).
 It excludes part of the compressed scenario 2) with $1 \leq \tau \leq 2$ (blue points), where VIB is particularly large (see Eq.~\eqref{eq:indirect}). The bound cannot directly be applied to scenarios with $\tau > 2$ (dark gray points), for which the photon energy peak is much less pronounced and the VIB search thus significantly loses sensitivity. Inclusive gamma-ray searches in nearby dwarf galaxies lead to a less strigent upper bound on $\langle \sigma_{\mu\bar\mu\gamma}v\rangle$ (black line), which we adopt from~\cite{GeringerSameth:2011iw} under the assumption that VIB dominates photon emission in DM annihilation. Parameter space regions with $\tau>2$ and a relic density comparable to the observed value lie at least an order of magnitude below the bound from dwarf galaxies. Notice that the Fermi bounds on the cross section are based on data from Pass 7. With the recently released Pass 8 data set~\cite{Ackermann:2015zua}, we expect these bounds to improve by about an order of magnitude. The inclusive gamma-ray search might thus start probing the heavy lepton scenario, as well as the part of the compressed scenario, which does not display a significant spectral feature.


\section{Conclusions and Outlook}\label{CO}
We have shown that a minimal phenomenological model featuring a new abelian force with radiative couplings to left-handed muons can simultaneously explain the observed muon anomalies in $g_{\mu}-2$ and $b\to s\mu^+\mu^-$ transitions through one-loop effects of new particles charged under the associated U(1)$_X$ symmetry. 
 The spontaneous breaking of U(1)$_X$ down to a $\mathbb{Z}_2$ symmetry induces a sizeable mass splitting among the scalar states of the dark sector, which controls the radiative $Z'$ coupling to muon pairs
 and ensures that the lightest new state is a cosmologically stable neutral scalar. Hence, addressing both the $g_\mu-2$ and $b\to s\mu^+\mu^-$
 anomalies in this way typically {\it yields} a leptophilic DM candidate, which is
 hardly visible in direct DM detection searches.
 
All muon-related anomalies can simultaneously be accommodated with $Z'$ and DM masses around the EW scale and large dark couplings $y,\,g'\gtrsim 2$. While DM annihilation is governed by the same interactions that explain the muon anomalies, the
large Yukawa
 coupling $y$ typically allows for a thermal DM relic density surprisingly close to the observed value, unless significant coannihilation is present.  
 This result partially relies on the fact that the dominant DM annihilation into lepton pairs is $d$-wave velocity-suppressed, thus favoring larger
 couplings compared to scenarios with $s-$ or $p-$wave annihilation. The velocity suppression furthermore depletes final-state and secondary photon emission from DM annihilation into muon pairs in our galactic neighorhood today. Signals in indirect detection experiments are therefore dominantly due to hard photons from VIB.
 
Another key ingredient of the model is a new vector-like lepton doublet, which in particular mediates tree-level DM annihilation to leptons. Being $\mathbb{Z}_2$-odd, these dark leptons do not mix with SM leptons, nor do they couple to the SM Higgs field at tree level.    
 This prevents the otherwise
 severe constraints from Higgs and EW precision observables, most notably oblique corrections. The only significant effect on low-energy observables lies in the $Z$ boson coupling to muon pairs or invisible states, which is 
  in mild tension with precision LEP data at the $Z$ pole.
  
At high-energy hadron colliders, dark leptons around the EW scale are produced in pairs through
Drell-Yan-like processes.
 We recast available LHC searches for muon pairs with missing energy
  in the context of supersymmetry and find a general lower bound on the dark lepton mass, $m_L \gtrsim 450\,$GeV. A good fit to all muon anomalies with a relic density around the observed value is still possible in two characteristic parameter regions. 
 These are 1) the {\it heavy lepton scenario} with dark leptons above $450\,$GeV and a DM candidate  in the range $125\lesssim m_{\chi_0}\lesssim 250\,$GeV, and 2) the {\it compressed scenario} with a light dark sector around $70\,$GeV and a small splitting between the dark-lepton and DM masses, $m_L-m_{\chi_0}\lesssim 60\,$GeV, to which current di-muon searches are insensitive.

 Accommodating the LHCb anomalies in the heavy lepton scenario
 typically requires a very large Yukawa coupling not far from the non-perturbative regime, which suggests a UV-completion of our model not far above the TeV scale. Weaker couplings are nevertheless possible in the compressed scenario
 thanks to the larger DM annihilation cross section at freeze-out in the limit of a small mass gap with the dark lepton. However, this small mass splitting induces a sharply peaked photon energy spectrum from VIB in DM annhihilation today. Dedicated searches for such a signature by the Fermi experiment partially
 exclude the compressed scenario as a simultaneous explanation of muon anomalies and thermal DM. Interestingly, indirect detection experiments are thus complementary to collider searches, which are not sensitive to the compressed scenario. Probing the heavy lepton scenario through indirect detection, however, is difficult, since the photon energy peak from VIB is much less pronounced for large mass gaps and inclusive gamma-ray searches are less sensitive.

We point out that $B_s$ meson mixing significantly limits the $Z'\bar b s$ coupling to at most a few permil.  Given the loop suppression of the $Z'$ coupling to muon pairs in our model, the $b\to s\mu^+\mu^-$ anomalies imply
 new contributions to $B_s$ meson mixing right around the corner, while they are only a possibility in most $Z'$ models explaining the LHCb anomalies with tree-level couplings to leptons. This is an interesting
  implication of simultaneously addressing the $g_{\mu}-2$ discrepancy. 
Moreover, the absence of a $Z'$ coupling to valence quarks and its constrained interaction with
 bottom and strange quarks strongly
  limits direct $Z'$ production at hadron colliders. In particular, resonant $Z'$ production followed by decays in muon pairs is well below the sensitivity of the 8TeV LHC. Similar conclusions apply to mono-jet searches.

Prospects are good that our model can be tested in the near future. Since dedicated DM searches through direct
 detection are not sensitive to our scenario, the 14$\,$TeV LHC will be the most important experiment to probe it.
 We expect that the reach for dark muons should be improved by close to a factor  of two during run II, similarly to what was found for smuons~\cite{Melzer-Pellmann:2014eta,Eckel:2014dza}. Thus, the parameter space 
of  the heavy lepton scenario
 will be fully covered, since the $g_{\mu}-2$ anomaly sets an upper bound $m_L\lesssim 800\,$GeV for couplings $y<4\pi$.
 The  spectrum in the compressed scenario (which is already under pressure from indirect DM detection searches)
 is more difficult to access through direct searches. One possibility would be to exploit the monojet signature from the direct production of dark muons in association with an ISR jet in the  highly
 compressed region, where the dark muon decay leads to an invisible signature.
 Precision measurements of $Z$ boson couplings at a future $e^+e^-$ collider would give  complementary indirect information about the dark sector. \\


\section{Acknowledgements}
We wish to thank Diego Guadagnoli, Kenneth Lane and Gilad Perez and Alexander Pukhov for discussions. The work of CD is supported by the ``Investissements d'avenir, Labex ENIGMASS''. SW is supported in part by the U.S. National Science Foundation under Grant PHY-1212635.\\

\appendix

\section{Auxiliary Functions}\label{App}
The $g_\mu-2$ loop function from the diagrams in FIG.~\ref{gm2diagram} is
\beq\label{Fg}
F_{g}(\tau)\equiv \frac{1}{6(\tau-1)^4}\left[\tau^3-6\tau(\tau-\log\tau)+3\tau+2\right]\,.
\eeq
For degenerate dark (pseudo-)scalar and dark lepton masses, one has $F_g(\tau=1)=1/12$, while in the decoupling limit of the dark lepton the function decreases as $F_g(\tau\gg1)\simeq 1/(6\tau)$.\\

For the dominant $Z'$ coupling to SM lepton pairs at zero-momentum, corresponding to the diagrams in FIG.~\ref{Zpmumu}, the loop function is
\beq\label{FZp}
F_{Z'}(\tau,\delta)&\equiv&  \int_0^1 du\int_0^{1-u}dv\log\left[1+\frac{ v\delta}{\tau+(u+v)(1-\tau)}\right] \nn\\
&&+\log\left[1-\frac{ v\delta}{\tau+(u+v)(1+\delta-\tau)}\right]\,.
\eeq
For degenerate scalar and pseudoscalar masses, $F_{Z'}(\tau,0)=0$, while in the limit of large mass splitting Eq.~\eqref{FZp} approximately gives 
\beq
F_{Z'}(\tau,\delta\gg 1)\simeq\frac{1}{2}\log\delta-\frac{\tau^2(1+\log\tau)-3\tau+2}{2(\tau-1)^2}\,.
\eeq
Moreover, Eq.~\eqref{FZp} approaches
 a constant when both mass splittings are comparably large, $F_{Z'}(\tau\sim\delta\gg1)\simeq1/4$.\\

The one-loop corrections to the $V=W,Z$ couplings to SM leptons from FIG.~\ref{gm2diagram} are proportional to the loop function
\begin{widetext}
\beq\label{FV}
F_V(\tau,r_q)=2\int_0^1 du\int_0^{1-u}dv\left[\frac{1-u-v}{u+v}\log f(\tau,u+v)
+\frac{1-f(\tau,u+v)+2uvr_q}{f(\tau,u+v)- u v r_q}-\log(f(\tau,u+v)-uvr_q)\right]\,,
\eeq
\end{widetext}
where $f(\tau,x)\equiv x+(1-x)/\tau$ and $r_q\equiv q^2/m_L^2$, with $q^2$ the squared momentum of the external gauge boson. The loop function falls
 off at low $q^2$ as
\beq
F_V(\tau,r_q\ll1)&\simeq& \frac{r_q \tau}{36(\tau-1)^4}\left[(\tau-1)(7\tau^2-29\tau+16)\right.\nn\\
&&\quad \left.+6(3\tau-2)\log\tau\right]. 
\eeq

Finally, the function controlling the approximate $3-$body correction  from VIB to the relic density is
\beq\label{3bodycorr}
F_\gamma(\tau)&=&(\tau +1)\left[\frac{\pi^2}{6}-\log^2\left(\frac{\tau+1}{2\tau}\right)-2{\rm Li}_2\left(\frac{\tau+1}{2\tau}\right)\right]\nonumber\\
&&+\frac{4\tau+3}{\tau+1}+\frac{4\tau^2-3\tau-1}{2\tau}\log\left(\frac{\tau-1}{\tau+1}\right)\,,
\eeq 
where ${\rm Li}_2(x)   = -\int_0^1dy\log(1-xy)/y$ is the polylogarithm function of second order.
\bibliographystyle{apsrev}
\bibliography{muonDM}

\begin{thebibliography}{56}
\expandafter\ifx\csname natexlab\endcsname\relax\def\natexlab#1{#1}\fi
\expandafter\ifx\csname bibnamefont\endcsname\relax
  \def\bibnamefont#1{#1}\fi
\expandafter\ifx\csname bibfnamefont\endcsname\relax
  \def\bibfnamefont#1{#1}\fi
\expandafter\ifx\csname citenamefont\endcsname\relax
  \def\citenamefont#1{#1}\fi
\expandafter\ifx\csname url\endcsname\relax
  \def\url#1{\texttt{#1}}\fi
\expandafter\ifx\csname urlprefix\endcsname\relax\def\urlprefix{URL }\fi
\providecommand{\bibinfo}[2]{#2}
\providecommand{\eprint}[2][]{\url{#2}}

\bibitem[{\citenamefont{Bennett et~al.}(2006)}]{gm2BNL}
\bibinfo{author}{\bibfnamefont{G.}~\bibnamefont{Bennett}} \bibnamefont{et~al.}
  (\bibinfo{collaboration}{Muon g-2}), \bibinfo{journal}{Phys.Rev.}
  \textbf{\bibinfo{volume}{D73}}, \bibinfo{pages}{072003}
  (\bibinfo{year}{2006}), \eprint{hep-ex/0602035}.

\bibitem[{\citenamefont{Aaij et~al.}(2013)}]{Kstar}
\bibinfo{author}{\bibfnamefont{R.}~\bibnamefont{Aaij}} \bibnamefont{et~al.}
  (\bibinfo{collaboration}{LHCb}), \bibinfo{journal}{Phys.Rev.Lett.}
  \textbf{\bibinfo{volume}{111}}, \bibinfo{pages}{191801}
  (\bibinfo{year}{2013}), \eprint{1308.1707}.

\bibitem[{Kst(2015)}]{Kstarnew}
 (\bibinfo{year}{2015}), \urlprefix\url{http://cds.cern.ch/record/2002772}.

\bibitem[{\citenamefont{Aaij et~al.}(2014)}]{LHCbRK}
\bibinfo{author}{\bibfnamefont{R.}~\bibnamefont{Aaij}} \bibnamefont{et~al.}
  (\bibinfo{collaboration}{LHCb}), \bibinfo{journal}{Phys.Rev.Lett.}
  \textbf{\bibinfo{volume}{113}}, \bibinfo{pages}{151601}
  (\bibinfo{year}{2014}), \eprint{1406.6482}.

\bibitem[{\citenamefont{Akerib et~al.}(2014)}]{Akerib:2013tjd}
\bibinfo{author}{\bibfnamefont{D.}~\bibnamefont{Akerib}} \bibnamefont{et~al.}
  (\bibinfo{collaboration}{LUX}), \bibinfo{journal}{Phys.Rev.Lett.}
  \textbf{\bibinfo{volume}{112}}, \bibinfo{pages}{091303}
  (\bibinfo{year}{2014}), \eprint{1310.8214}.

\bibitem[{\citenamefont{Olive et~al.}(2014)}]{PDG}
\bibinfo{author}{\bibfnamefont{K.}~\bibnamefont{Olive}} \bibnamefont{et~al.}
  (\bibinfo{collaboration}{Particle Data Group}), \bibinfo{journal}{Chin.Phys.}
  \textbf{\bibinfo{volume}{C38}}, \bibinfo{pages}{090001}
  (\bibinfo{year}{2014}).

\bibitem[{\citenamefont{Khachatryan et~al.}(2015{\natexlab{a}})}]{CMS:2014xfa}
\bibinfo{author}{\bibfnamefont{V.}~\bibnamefont{Khachatryan}}
  \bibnamefont{et~al.} (\bibinfo{collaboration}{CMS, LHCb}),
  \bibinfo{journal}{Nature} \textbf{\bibinfo{volume}{522}}, \bibinfo{pages}{68}
  (\bibinfo{year}{2015}{\natexlab{a}}), \eprint{1411.4413}.

\bibitem[{\citenamefont{Descotes-Genon
  et~al.}(2013)\citenamefont{Descotes-Genon, Matias, and
  Virto}}]{Descotes-Genon:2013wba}
\bibinfo{author}{\bibfnamefont{S.}~\bibnamefont{Descotes-Genon}},
  \bibinfo{author}{\bibfnamefont{J.}~\bibnamefont{Matias}}, \bibnamefont{and}
  \bibinfo{author}{\bibfnamefont{J.}~\bibnamefont{Virto}},
  \bibinfo{journal}{Phys.Rev.} \textbf{\bibinfo{volume}{D88}},
  \bibinfo{pages}{074002} (\bibinfo{year}{2013}), \eprint{1307.5683}.

\bibitem[{\citenamefont{Altmannshofer and Straub}(2013)}]{Altmannshofer-Straub}
\bibinfo{author}{\bibfnamefont{W.}~\bibnamefont{Altmannshofer}}
  \bibnamefont{and} \bibinfo{author}{\bibfnamefont{D.~M.}
  \bibnamefont{Straub}}, \bibinfo{journal}{Eur.Phys.J.}
  \textbf{\bibinfo{volume}{C73}}, \bibinfo{pages}{2646} (\bibinfo{year}{2013}),
  \eprint{1308.1501}.

\bibitem[{\citenamefont{Beaujean et~al.}(2014)\citenamefont{Beaujean, Bobeth,
  and van Dyk}}]{Beaujean:2013soa}
\bibinfo{author}{\bibfnamefont{F.}~\bibnamefont{Beaujean}},
  \bibinfo{author}{\bibfnamefont{C.}~\bibnamefont{Bobeth}}, \bibnamefont{and}
  \bibinfo{author}{\bibfnamefont{D.}~\bibnamefont{van Dyk}},
  \bibinfo{journal}{Eur. Phys. J.} \textbf{\bibinfo{volume}{C74}},
  \bibinfo{pages}{2897} (\bibinfo{year}{2014}), \bibinfo{note}{[Erratum: Eur.
  Phys. J.C74,3179(2014)]}, \eprint{1310.2478}.

\bibitem[{\citenamefont{Altmannshofer and
  Straub}(2014)}]{Altmannshofer:2014rta}
\bibinfo{author}{\bibfnamefont{W.}~\bibnamefont{Altmannshofer}}
  \bibnamefont{and} \bibinfo{author}{\bibfnamefont{D.~M.} \bibnamefont{Straub}}
  (\bibinfo{year}{2014}), \eprint{1411.3161}.

\bibitem[{\citenamefont{Hurth et~al.}(2014)\citenamefont{Hurth, Mahmoudi, and
  Neshatpour}}]{Hurth:2014vma}
\bibinfo{author}{\bibfnamefont{T.}~\bibnamefont{Hurth}},
  \bibinfo{author}{\bibfnamefont{F.}~\bibnamefont{Mahmoudi}}, \bibnamefont{and}
  \bibinfo{author}{\bibfnamefont{S.}~\bibnamefont{Neshatpour}},
  \bibinfo{journal}{JHEP} \textbf{\bibinfo{volume}{1412}}, \bibinfo{pages}{053}
  (\bibinfo{year}{2014}), \eprint{1410.4545}.

\bibitem[{\citenamefont{Gauld et~al.}(2014)\citenamefont{Gauld, Goertz, and
  Haisch}}]{Gauld:2013qba}
\bibinfo{author}{\bibfnamefont{R.}~\bibnamefont{Gauld}},
  \bibinfo{author}{\bibfnamefont{F.}~\bibnamefont{Goertz}}, \bibnamefont{and}
  \bibinfo{author}{\bibfnamefont{U.}~\bibnamefont{Haisch}},
  \bibinfo{journal}{Phys.Rev.} \textbf{\bibinfo{volume}{D89}},
  \bibinfo{pages}{015005} (\bibinfo{year}{2014}), \eprint{1308.1959}.

\bibitem[{\citenamefont{Altmannshofer et~al.}(2014)\citenamefont{Altmannshofer,
  Gori, Pospelov, and Yavin}}]{Altmannshofer:2014cfa}
\bibinfo{author}{\bibfnamefont{W.}~\bibnamefont{Altmannshofer}},
  \bibinfo{author}{\bibfnamefont{S.}~\bibnamefont{Gori}},
  \bibinfo{author}{\bibfnamefont{M.}~\bibnamefont{Pospelov}}, \bibnamefont{and}
  \bibinfo{author}{\bibfnamefont{I.}~\bibnamefont{Yavin}},
  \bibinfo{journal}{Phys.Rev.} \textbf{\bibinfo{volume}{D89}},
  \bibinfo{pages}{095033} (\bibinfo{year}{2014}), \eprint{1403.1269}.

\bibitem[{\citenamefont{Aristizabal~Sierra
  et~al.}(2015)\citenamefont{Aristizabal~Sierra, Staub, and
  Vicente}}]{Sierra:2015fma}
\bibinfo{author}{\bibfnamefont{D.}~\bibnamefont{Aristizabal~Sierra}},
  \bibinfo{author}{\bibfnamefont{F.}~\bibnamefont{Staub}}, \bibnamefont{and}
  \bibinfo{author}{\bibfnamefont{A.}~\bibnamefont{Vicente}},
  \bibinfo{journal}{Phys. Rev.} \textbf{\bibinfo{volume}{D92}},
  \bibinfo{pages}{015001} (\bibinfo{year}{2015}), \eprint{1503.06077}.

\bibitem[{\citenamefont{Hiller and Schmaltz}(2014)}]{Hiller:2014yaa}
\bibinfo{author}{\bibfnamefont{G.}~\bibnamefont{Hiller}} \bibnamefont{and}
  \bibinfo{author}{\bibfnamefont{M.}~\bibnamefont{Schmaltz}},
  \bibinfo{journal}{Phys.Rev.} \textbf{\bibinfo{volume}{D90}},
  \bibinfo{pages}{054014} (\bibinfo{year}{2014}), \eprint{1408.1627}.

\bibitem[{\citenamefont{Biswas et~al.}(2015)\citenamefont{Biswas, Chowdhury,
  Han, and Lee}}]{Biswas:2014gga}
\bibinfo{author}{\bibfnamefont{S.}~\bibnamefont{Biswas}},
  \bibinfo{author}{\bibfnamefont{D.}~\bibnamefont{Chowdhury}},
  \bibinfo{author}{\bibfnamefont{S.}~\bibnamefont{Han}}, \bibnamefont{and}
  \bibinfo{author}{\bibfnamefont{S.~J.} \bibnamefont{Lee}},
  \bibinfo{journal}{JHEP} \textbf{\bibinfo{volume}{1502}}, \bibinfo{pages}{142}
  (\bibinfo{year}{2015}), \eprint{1409.0882}.

\bibitem[{\citenamefont{Gripaios et~al.}(2015)\citenamefont{Gripaios,
  Nardecchia, and Renner}}]{Gripaios:2014tna}
\bibinfo{author}{\bibfnamefont{B.}~\bibnamefont{Gripaios}},
  \bibinfo{author}{\bibfnamefont{M.}~\bibnamefont{Nardecchia}},
  \bibnamefont{and} \bibinfo{author}{\bibfnamefont{S.~A.}
  \bibnamefont{Renner}}, \bibinfo{journal}{JHEP} \textbf{\bibinfo{volume}{05}},
  \bibinfo{pages}{006} (\bibinfo{year}{2015}), \eprint{1412.1791}.

\bibitem[{\citenamefont{Fox et~al.}(2011)\citenamefont{Fox, Liu, Tucker-Smith,
  and Weiner}}]{effectiveZprime}
\bibinfo{author}{\bibfnamefont{P.~J.} \bibnamefont{Fox}},
  \bibinfo{author}{\bibfnamefont{J.}~\bibnamefont{Liu}},
  \bibinfo{author}{\bibfnamefont{D.}~\bibnamefont{Tucker-Smith}},
  \bibnamefont{and} \bibinfo{author}{\bibfnamefont{N.}~\bibnamefont{Weiner}},
  \bibinfo{journal}{Phys.Rev.} \textbf{\bibinfo{volume}{D84}},
  \bibinfo{pages}{115006} (\bibinfo{year}{2011}), \eprint{1104.4127}.

\bibitem[{\citenamefont{Kumar and Wells}(2006)}]{Kumar:2006gm}
\bibinfo{author}{\bibfnamefont{J.}~\bibnamefont{Kumar}} \bibnamefont{and}
  \bibinfo{author}{\bibfnamefont{J.~D.} \bibnamefont{Wells}},
  \bibinfo{journal}{Phys. Rev.} \textbf{\bibinfo{volume}{D74}},
  \bibinfo{pages}{115017} (\bibinfo{year}{2006}), \eprint{hep-ph/0606183}.

\bibitem[{\citenamefont{Chun et~al.}(2011)\citenamefont{Chun, Park, and
  Scopel}}]{Chun:2010ve}
\bibinfo{author}{\bibfnamefont{E.~J.} \bibnamefont{Chun}},
  \bibinfo{author}{\bibfnamefont{J.-C.} \bibnamefont{Park}}, \bibnamefont{and}
  \bibinfo{author}{\bibfnamefont{S.}~\bibnamefont{Scopel}},
  \bibinfo{journal}{JHEP} \textbf{\bibinfo{volume}{02}}, \bibinfo{pages}{100}
  (\bibinfo{year}{2011}), \eprint{1011.3300}.

\bibitem[{\citenamefont{Djouadi et~al.}(2012)\citenamefont{Djouadi, Lebedev,
  Mambrini, and Quevillon}}]{Djouadi:2011aa}
\bibinfo{author}{\bibfnamefont{A.}~\bibnamefont{Djouadi}},
  \bibinfo{author}{\bibfnamefont{O.}~\bibnamefont{Lebedev}},
  \bibinfo{author}{\bibfnamefont{Y.}~\bibnamefont{Mambrini}}, \bibnamefont{and}
  \bibinfo{author}{\bibfnamefont{J.}~\bibnamefont{Quevillon}},
  \bibinfo{journal}{Phys. Lett.} \textbf{\bibinfo{volume}{B709}},
  \bibinfo{pages}{65} (\bibinfo{year}{2012}), \eprint{1112.3299}.

\bibitem[{\citenamefont{Buras et~al.}(2013)\citenamefont{Buras, De~Fazio, and
  Girrbach}}]{Buras-DeFazio-Girrbach}
\bibinfo{author}{\bibfnamefont{A.~J.} \bibnamefont{Buras}},
  \bibinfo{author}{\bibfnamefont{F.}~\bibnamefont{De~Fazio}}, \bibnamefont{and}
  \bibinfo{author}{\bibfnamefont{J.}~\bibnamefont{Girrbach}},
  \bibinfo{journal}{JHEP} \textbf{\bibinfo{volume}{1302}}, \bibinfo{pages}{116}
  (\bibinfo{year}{2013}), \eprint{1211.1896}.

\bibitem[{\citenamefont{Glashow et~al.}(2015)\citenamefont{Glashow, Guadagnoli,
  and Lane}}]{Glashow:2014iga}
\bibinfo{author}{\bibfnamefont{S.~L.} \bibnamefont{Glashow}},
  \bibinfo{author}{\bibfnamefont{D.}~\bibnamefont{Guadagnoli}},
  \bibnamefont{and} \bibinfo{author}{\bibfnamefont{K.}~\bibnamefont{Lane}},
  \bibinfo{journal}{Phys.Rev.Lett.} \textbf{\bibinfo{volume}{114}},
  \bibinfo{pages}{091801} (\bibinfo{year}{2015}), \eprint{1411.0565}.

\bibitem[{\citenamefont{Crivellin
  et~al.}(2015{\natexlab{a}})\citenamefont{Crivellin, D'Ambrosio, and
  Heeck}}]{Crivellin1}
\bibinfo{author}{\bibfnamefont{A.}~\bibnamefont{Crivellin}},
  \bibinfo{author}{\bibfnamefont{G.}~\bibnamefont{D'Ambrosio}},
  \bibnamefont{and} \bibinfo{author}{\bibfnamefont{J.}~\bibnamefont{Heeck}},
  \bibinfo{journal}{Phys.Rev.Lett.} \textbf{\bibinfo{volume}{114}},
  \bibinfo{pages}{151801} (\bibinfo{year}{2015}{\natexlab{a}}),
  \eprint{1501.00993}.

\bibitem[{\citenamefont{Crivellin
  et~al.}(2015{\natexlab{b}})\citenamefont{Crivellin, D'Ambrosio, and
  Heeck}}]{Crivellin2}
\bibinfo{author}{\bibfnamefont{A.}~\bibnamefont{Crivellin}},
  \bibinfo{author}{\bibfnamefont{G.}~\bibnamefont{D'Ambrosio}},
  \bibnamefont{and} \bibinfo{author}{\bibfnamefont{J.}~\bibnamefont{Heeck}},
  \bibinfo{journal}{Phys.Rev.} \textbf{\bibinfo{volume}{D91}},
  \bibinfo{pages}{075006} (\bibinfo{year}{2015}{\natexlab{b}}),
  \eprint{1503.03477}.

\bibitem[{\citenamefont{Lee and Tandean}(2015)}]{MLFVbs}
\bibinfo{author}{\bibfnamefont{C.-J.} \bibnamefont{Lee}} \bibnamefont{and}
  \bibinfo{author}{\bibfnamefont{J.}~\bibnamefont{Tandean}}
  (\bibinfo{year}{2015}), \eprint{1505.04692}.

\bibitem[{\citenamefont{Altmannshofer and
  Straub}(2015)}]{Altmannshofer:2015sma}
\bibinfo{author}{\bibfnamefont{W.}~\bibnamefont{Altmannshofer}}
  \bibnamefont{and} \bibinfo{author}{\bibfnamefont{D.~M.} \bibnamefont{Straub}}
  (\bibinfo{year}{2015}), \eprint{1503.06199}.

\bibitem[{\citenamefont{Aad et~al.}(2014{\natexlab{a}})}]{Aad:2014cka}
\bibinfo{author}{\bibfnamefont{G.}~\bibnamefont{Aad}} \bibnamefont{et~al.}
  (\bibinfo{collaboration}{ATLAS}), \bibinfo{journal}{Phys.Rev.}
  \textbf{\bibinfo{volume}{D90}}, \bibinfo{pages}{052005}
  (\bibinfo{year}{2014}{\natexlab{a}}), \eprint{1405.4123}.

\bibitem[{\citenamefont{Khachatryan
  et~al.}(2015{\natexlab{b}})}]{Khachatryan:2014fba}
\bibinfo{author}{\bibfnamefont{V.}~\bibnamefont{Khachatryan}}
  \bibnamefont{et~al.} (\bibinfo{collaboration}{CMS}), \bibinfo{journal}{JHEP}
  \textbf{\bibinfo{volume}{1504}}, \bibinfo{pages}{025}
  (\bibinfo{year}{2015}{\natexlab{b}}), \eprint{1412.6302}.

\bibitem[{\citenamefont{Aad et~al.}(2015)}]{Aad:2015zva}
\bibinfo{author}{\bibfnamefont{G.}~\bibnamefont{Aad}} \bibnamefont{et~al.}
  (\bibinfo{collaboration}{ATLAS}), \bibinfo{journal}{Eur.Phys.J.}
  \textbf{\bibinfo{volume}{C75}}, \bibinfo{pages}{299} (\bibinfo{year}{2015}),
  \eprint{1502.01518}.

\bibitem[{\citenamefont{Khachatryan
  et~al.}(2015{\natexlab{c}})}]{Khachatryan:2014rra}
\bibinfo{author}{\bibfnamefont{V.}~\bibnamefont{Khachatryan}}
  \bibnamefont{et~al.} (\bibinfo{collaboration}{CMS}),
  \bibinfo{journal}{Eur.Phys.J.} \textbf{\bibinfo{volume}{C75}},
  \bibinfo{pages}{235} (\bibinfo{year}{2015}{\natexlab{c}}),
  \eprint{1408.3583}.

\bibitem[{\citenamefont{Arina et~al.}(2015)\citenamefont{Arina, Catalan, Kraml,
  Kulkarni, and Laa}}]{Arina:2015uea}
\bibinfo{author}{\bibfnamefont{C.}~\bibnamefont{Arina}},
  \bibinfo{author}{\bibfnamefont{M.~E.~C.} \bibnamefont{Catalan}},
  \bibinfo{author}{\bibfnamefont{S.}~\bibnamefont{Kraml}},
  \bibinfo{author}{\bibfnamefont{S.}~\bibnamefont{Kulkarni}}, \bibnamefont{and}
  \bibinfo{author}{\bibfnamefont{U.}~\bibnamefont{Laa}},
  \bibinfo{journal}{JHEP} \textbf{\bibinfo{volume}{1505}}, \bibinfo{pages}{142}
  (\bibinfo{year}{2015}), \eprint{1503.02960}.

\bibitem[{\citenamefont{Kraml et~al.}(2014)\citenamefont{Kraml, Kulkarni, Laa,
  Lessa, Magerl et~al.}}]{Kraml:2014sna}
\bibinfo{author}{\bibfnamefont{S.}~\bibnamefont{Kraml}},
  \bibinfo{author}{\bibfnamefont{S.}~\bibnamefont{Kulkarni}},
  \bibinfo{author}{\bibfnamefont{U.}~\bibnamefont{Laa}},
  \bibinfo{author}{\bibfnamefont{A.}~\bibnamefont{Lessa}},
  \bibinfo{author}{\bibfnamefont{V.}~\bibnamefont{Magerl}},
  \bibnamefont{et~al.} (\bibinfo{year}{2014}), \eprint{1412.1745}.

\bibitem[{\citenamefont{Aad et~al.}(2014{\natexlab{b}})}]{Aad:2014vma}
\bibinfo{author}{\bibfnamefont{G.}~\bibnamefont{Aad}} \bibnamefont{et~al.}
  (\bibinfo{collaboration}{ATLAS}), \bibinfo{journal}{JHEP}
  \textbf{\bibinfo{volume}{05}}, \bibinfo{pages}{071}
  (\bibinfo{year}{2014}{\natexlab{b}}), \eprint{1403.5294}.

\bibitem[{\citenamefont{Khachatryan et~al.}(2014)}]{Khachatryan:2014qwa}
\bibinfo{author}{\bibfnamefont{V.}~\bibnamefont{Khachatryan}}
  \bibnamefont{et~al.} (\bibinfo{collaboration}{CMS}), \bibinfo{journal}{Eur.
  Phys. J.} \textbf{\bibinfo{volume}{C74}}, \bibinfo{pages}{3036}
  (\bibinfo{year}{2014}), \eprint{1405.7570}.

\bibitem[{\citenamefont{Abdallah et~al.}(2003)}]{Abdallah:2003xe}
\bibinfo{author}{\bibfnamefont{J.}~\bibnamefont{Abdallah}} \bibnamefont{et~al.}
  (\bibinfo{collaboration}{DELPHI}), \bibinfo{journal}{Eur. Phys. J.}
  \textbf{\bibinfo{volume}{C31}}, \bibinfo{pages}{421} (\bibinfo{year}{2003}),
  \eprint{hep-ex/0311019}.

\bibitem[{\citenamefont{Giacchino et~al.}(2013)\citenamefont{Giacchino,
  Lopez-Honorez, and Tytgat}}]{ScalarDMTytgat}
\bibinfo{author}{\bibfnamefont{F.}~\bibnamefont{Giacchino}},
  \bibinfo{author}{\bibfnamefont{L.}~\bibnamefont{Lopez-Honorez}},
  \bibnamefont{and} \bibinfo{author}{\bibfnamefont{M.~H.}
  \bibnamefont{Tytgat}}, \bibinfo{journal}{JCAP}
  \textbf{\bibinfo{volume}{1310}}, \bibinfo{pages}{025} (\bibinfo{year}{2013}),
  \eprint{1307.6480}.

\bibitem[{\citenamefont{Gondolo and Gelmini}(1991)}]{GelminiGondolo}
\bibinfo{author}{\bibfnamefont{P.}~\bibnamefont{Gondolo}} \bibnamefont{and}
  \bibinfo{author}{\bibfnamefont{G.}~\bibnamefont{Gelmini}},
  \bibinfo{journal}{Nucl.Phys.} \textbf{\bibinfo{volume}{B360}},
  \bibinfo{pages}{145} (\bibinfo{year}{1991}).

\bibitem[{\citenamefont{Ade et~al.}(2015)}]{Ade:2015xua}
\bibinfo{author}{\bibfnamefont{P.}~\bibnamefont{Ade}} \bibnamefont{et~al.}
  (\bibinfo{collaboration}{Planck}) (\bibinfo{year}{2015}),
  \eprint{1502.01589}.

\bibitem[{\citenamefont{Bringmann et~al.}(2008)\citenamefont{Bringmann,
  Bergstrom, and Edsjo}}]{Bringmann:2007nk}
\bibinfo{author}{\bibfnamefont{T.}~\bibnamefont{Bringmann}},
  \bibinfo{author}{\bibfnamefont{L.}~\bibnamefont{Bergstrom}},
  \bibnamefont{and} \bibinfo{author}{\bibfnamefont{J.}~\bibnamefont{Edsjo}},
  \bibinfo{journal}{JHEP} \textbf{\bibinfo{volume}{01}}, \bibinfo{pages}{049}
  (\bibinfo{year}{2008}), \eprint{0710.3169}.

\bibitem[{\citenamefont{Toma}(2013)}]{Toma:2013bka}
\bibinfo{author}{\bibfnamefont{T.}~\bibnamefont{Toma}},
  \bibinfo{journal}{Phys.Rev.Lett.} \textbf{\bibinfo{volume}{111}},
  \bibinfo{pages}{091301} (\bibinfo{year}{2013}), \eprint{1307.6181}.

\bibitem[{\citenamefont{B\'elanger et~al.}(2015)\citenamefont{B\'elanger,
  Boudjema, Pukhov, and Semenov}}]{MO41}
\bibinfo{author}{\bibfnamefont{G.}~\bibnamefont{B\'elanger}},
  \bibinfo{author}{\bibfnamefont{F.}~\bibnamefont{Boudjema}},
  \bibinfo{author}{\bibfnamefont{A.}~\bibnamefont{Pukhov}}, \bibnamefont{and}
  \bibinfo{author}{\bibfnamefont{A.}~\bibnamefont{Semenov}},
  \bibinfo{journal}{Comput.Phys.Commun.} \textbf{\bibinfo{volume}{192}},
  \bibinfo{pages}{322} (\bibinfo{year}{2015}), \eprint{1407.6129}.

\bibitem[{\citenamefont{Bringmann et~al.}(2012)\citenamefont{Bringmann, Huang,
  Ibarra, Vogl, and Weniger}}]{Bringmann:2012vr}
\bibinfo{author}{\bibfnamefont{T.}~\bibnamefont{Bringmann}},
  \bibinfo{author}{\bibfnamefont{X.}~\bibnamefont{Huang}},
  \bibinfo{author}{\bibfnamefont{A.}~\bibnamefont{Ibarra}},
  \bibinfo{author}{\bibfnamefont{S.}~\bibnamefont{Vogl}}, \bibnamefont{and}
  \bibinfo{author}{\bibfnamefont{C.}~\bibnamefont{Weniger}},
  \bibinfo{journal}{JCAP} \textbf{\bibinfo{volume}{1207}}, \bibinfo{pages}{054}
  (\bibinfo{year}{2012}), \eprint{1203.1312}.

\bibitem[{\citenamefont{Ackermann et~al.}(2011)}]{Ackermann:2011wa}
\bibinfo{author}{\bibfnamefont{M.}~\bibnamefont{Ackermann}}
  \bibnamefont{et~al.} (\bibinfo{collaboration}{Fermi-LAT}),
  \bibinfo{journal}{Phys. Rev. Lett.} \textbf{\bibinfo{volume}{107}},
  \bibinfo{pages}{241302} (\bibinfo{year}{2011}), \eprint{1108.3546}.

\bibitem[{\citenamefont{Ackermann et~al.}(2015)}]{Ackermann:2015zua}
\bibinfo{author}{\bibfnamefont{M.}~\bibnamefont{Ackermann}}
  \bibnamefont{et~al.} (\bibinfo{collaboration}{Fermi-LAT})
  (\bibinfo{year}{2015}), \eprint{1503.02641}.

\bibitem[{\citenamefont{Geringer-Sameth and
  Koushiappas}(2011)}]{GeringerSameth:2011iw}
\bibinfo{author}{\bibfnamefont{A.}~\bibnamefont{Geringer-Sameth}}
  \bibnamefont{and} \bibinfo{author}{\bibfnamefont{S.~M.}
  \bibnamefont{Koushiappas}}, \bibinfo{journal}{Phys. Rev. Lett.}
  \textbf{\bibinfo{volume}{107}}, \bibinfo{pages}{241303}
  (\bibinfo{year}{2011}), \eprint{1108.2914}.

\bibitem[{\citenamefont{Melzer-Pellmann and
  Pralavorio}(2014)}]{Melzer-Pellmann:2014eta}
\bibinfo{author}{\bibfnamefont{I.}~\bibnamefont{Melzer-Pellmann}}
  \bibnamefont{and}
  \bibinfo{author}{\bibfnamefont{P.}~\bibnamefont{Pralavorio}},
  \bibinfo{journal}{Eur.Phys.J.} \textbf{\bibinfo{volume}{C74}},
  \bibinfo{pages}{2801} (\bibinfo{year}{2014}), \eprint{1404.7191}.

\bibitem[{\citenamefont{Eckel et~al.}(2014)\citenamefont{Eckel, Ramsey-Musolf,
  Shepherd, and Su}}]{Eckel:2014dza}
\bibinfo{author}{\bibfnamefont{J.}~\bibnamefont{Eckel}},
  \bibinfo{author}{\bibfnamefont{M.~J.} \bibnamefont{Ramsey-Musolf}},
  \bibinfo{author}{\bibfnamefont{W.}~\bibnamefont{Shepherd}}, \bibnamefont{and}
  \bibinfo{author}{\bibfnamefont{S.}~\bibnamefont{Su}}, \bibinfo{journal}{JHEP}
  \textbf{\bibinfo{volume}{1411}}, \bibinfo{pages}{117} (\bibinfo{year}{2014}),
  \eprint{1408.2841}.

\bibitem[{\citenamefont{Pohl et~al.}(2010)}]{muHnature}
\bibinfo{author}{\bibfnamefont{R.}~\bibnamefont{Pohl}} \bibnamefont{et~al.},
  \bibinfo{journal}{Nature} \textbf{\bibinfo{volume}{466}},
  \bibinfo{pages}{213} (\bibinfo{year}{2010}).

\bibitem[{\citenamefont{Barger et~al.}(2011)\citenamefont{Barger, Chiang,
  Keung, and Marfatia}}]{psizeBarger}
\bibinfo{author}{\bibfnamefont{V.}~\bibnamefont{Barger}},
  \bibinfo{author}{\bibfnamefont{C.-W.} \bibnamefont{Chiang}},
  \bibinfo{author}{\bibfnamefont{W.-Y.} \bibnamefont{Keung}}, \bibnamefont{and}
  \bibinfo{author}{\bibfnamefont{D.}~\bibnamefont{Marfatia}},
  \bibinfo{journal}{Phys. Rev. Lett.} \textbf{\bibinfo{volume}{106}},
  \bibinfo{pages}{153001} (\bibinfo{year}{2011}), \eprint{1011.3519}.

\bibitem[{\citenamefont{Tucker-Smith and Yavin}(2011)}]{MeVforces}
\bibinfo{author}{\bibfnamefont{D.}~\bibnamefont{Tucker-Smith}}
  \bibnamefont{and} \bibinfo{author}{\bibfnamefont{I.}~\bibnamefont{Yavin}},
  \bibinfo{journal}{Phys. Rev.} \textbf{\bibinfo{volume}{D83}},
  \bibinfo{pages}{101702} (\bibinfo{year}{2011}), \eprint{1011.4922}.

\bibitem[{\citenamefont{Holdom}(1986)}]{HoldomU1}
\bibinfo{author}{\bibfnamefont{B.}~\bibnamefont{Holdom}},
  \bibinfo{journal}{Phys.Lett.} \textbf{\bibinfo{volume}{B166}},
  \bibinfo{pages}{196} (\bibinfo{year}{1986}).

\bibitem[{\citenamefont{Peskin and Takeuchi}(1990)}]{PeskinSTU}
\bibinfo{author}{\bibfnamefont{M.~E.} \bibnamefont{Peskin}} \bibnamefont{and}
  \bibinfo{author}{\bibfnamefont{T.}~\bibnamefont{Takeuchi}},
  \bibinfo{journal}{Phys.Rev.Lett.} \textbf{\bibinfo{volume}{65}},
  \bibinfo{pages}{964} (\bibinfo{year}{1990}).

\bibitem[{\citenamefont{Belyaev et~al.}(2013)\citenamefont{Belyaev,
  Christensen, and Pukhov}}]{CalcHEP34}
\bibinfo{author}{\bibfnamefont{A.}~\bibnamefont{Belyaev}},
  \bibinfo{author}{\bibfnamefont{N.~D.} \bibnamefont{Christensen}},
  \bibnamefont{and} \bibinfo{author}{\bibfnamefont{A.}~\bibnamefont{Pukhov}},
  \bibinfo{journal}{Comput.Phys.Commun.} \textbf{\bibinfo{volume}{184}},
  \bibinfo{pages}{1729} (\bibinfo{year}{2013}), \eprint{1207.6082}.

\bibitem[{\citenamefont{Billard et~al.}(2014)\citenamefont{Billard, Strigari,
  and Figueroa-Feliciano}}]{BillardCNB}
\bibinfo{author}{\bibfnamefont{J.}~\bibnamefont{Billard}},
  \bibinfo{author}{\bibfnamefont{L.}~\bibnamefont{Strigari}}, \bibnamefont{and}
  \bibinfo{author}{\bibfnamefont{E.}~\bibnamefont{Figueroa-Feliciano}},
  \bibinfo{journal}{Phys.Rev.} \textbf{\bibinfo{volume}{D89}},
  \bibinfo{pages}{023524} (\bibinfo{year}{2014}), \eprint{1307.5458}.

\end{thebibliography}

\end{document}